\documentclass[]{aa} 

\usepackage{natbib}
\bibpunct{(}{)}{;}{a}{}{,} 
\usepackage{epsfig}

\newcommand{\beq}{\begin{equation}} 
\newcommand{\eeq}{\end{equation}} 
\newcommand{\beqa}{\begin{eqnarray}} 
\newcommand{\eeqa}{\end{eqnarray}} 
\newcommand{\bea}{\begin{array}} 
\newcommand{\ea}{\end{array}} 

\newcommand{\dd}{{\rm d}}

\newcommand{\lag}{\langle} 
\newcommand{\rag}{\rangle} 

\newcommand{\ii}{{\rm i}}

\newcommand{\bb}{\overline{b}}
\newcommand{\nb}{\overline{n}}

\newcommand{\xih}{\xi^{\rm h}}
\newcommand{\xis}{\xi^{(s)}}
\newcommand{\xihs}{\xi^{{\rm h}(s)}}
\newcommand{\zetah}{\zeta^{\rm h}}
\newcommand{\etah}{\eta^{\rm h}}

\newcommand{\ve}{{\bf e}}
\newcommand{\vk}{{\bf k}}
\newcommand{\vOm}{{\bf \Omega}}

\newcommand{\vr}{{\bf r}}
\newcommand{\vs}{{\bf s}}
\newcommand{\vtheta}{{\vec{\theta}}}
\newcommand{\vv}{{\bf v}}

\newcommand{\vx}{{\bf x}}
\newcommand{\vy}{{\bf y}}

\newcommand{\tdelta}{{\tilde{\delta}}}

\newcommand{\tvv}{\tilde{\bf v}}
\newcommand{\tW}{\tilde{W}}

\newcommand{\hn}{\hat{n}}

\newcommand{\hW}{\hat{W}}

\newcommand{\hxi}{\hat{\xi}}

\newcommand{\cD}{{\cal D}}

\newcommand{\cI}{{\cal I}}

\newcommand{\cR}{{\cal R}}
\newcommand{\cV}{{\cal V}}
\newcommand{\cW}{{\cal W}}

\newcommand{\xicyl}{\overline{\xi_{\rm cyl}}}

\newcommand{\QQ}{Q}

\newcommand{\Rim}{R_{i,-}}
\newcommand{\Rip}{R_{i,+}}
\newcommand{\Rjm}{R_{j,-}}
\newcommand{\Rjp}{R_{j,+}}

\newcommand{\degs}{deg$^{2}$}
\newcommand{\flux}{$\rm erg \, s^{-1} \, cm^{-2}$}

\begin{document}

\title{Redshift-space correlation functions in large galaxy cluster surveys}    


\author{
P. Valageas\inst{1}
\and
N. Clerc\inst{2}
}
\institute{
Institut de Physique Th\'eorique, CEA Saclay, 91191 Gif-sur-Yvette, France
\and
Max-Planck-Institut f\"ur extraterrestrische Physik (MPE), Giessenbachstrasse 1, 85748 Garching, Germany
}  
\date{Received / Accepted } 
 
\abstract
{Large ongoing and upcoming galaxy cluster surveys in the optical, X-ray and millimetric wavelengths will provide rich samples of galaxy clusters at unprecedented depths. One key observable for constraining cosmological models is the correlation function of these objects, measured through their spectroscopic redshift.}
{We study the redshift-space correlation functions of clusters of galaxies,
averaged over finite redshift intervals, and their covariance matrices.
Expanding as usual the angular anisotropy of the redshift-space correlation
on Legendre polynomials, we consider the redshift-space distortions of the 
monopole as well as the next two multipoles, $2\ell=2$ and $4$.}
{Taking into account the Kaiser effect, we developed an analytical formalism to
obtain explicit expressions of all contributions to these mean correlations and covariance matrices.
We include shot-noise and sample-variance effects as well as Gaussian
and non-Gaussian contributions.}
{We obtain a reasonable agreement with numerical simulations for the mean
correlations and covariance matrices on large scales ($r> 10 h^{-1}$Mpc).
Redshift-space distortions amplify the monopole correlation by about
$10-20\%$, depending on the halo mass, but the signal-to-noise ratio remains of 
the same order as for the real-space correlation. This distortion will be significant
for surveys such as DES, Erosita, and Euclid, which should also measure the
quadrupole $2\ell=2$. The third multipole, $2\ell=4$, may only be marginally
detected by Euclid.}
{}

\keywords{Surveys - Galaxies: clusters: general - Cosmology: large-scale structure of Universe - Cosmology: observations}

\maketitle

\section{Introduction} 
\label{Introduction}

The growth of structure in our Universe results from a competition between gravitational attraction and the mean Hubble expansion. Studying recent structures provides an understanding of both these phenomena along with supplementary insights into the initial conditions that describe primordial density fluctuations, which are assumed to be nearly Gaussian-distributed. Analyses of massive large galaxy surveys \citep{Cole2005, Moles2008, Eisenstein2011, Blake2011}, galaxy cluster surveys \citep{Bohringer2001, Burenin2007, Koester2007a, Vanderlinde2010, Mehrtens2011}, weak-lensing studies \citep[e.g.][]{Benjamin2007, Massey2007, Kilbinger2009, Schrabback2010}, and measurements of the baryonic acoustic oscillation peak \citep{Eisenstein2005, Blake2011BAO, Slosar2011} are currently under way or are planned to probe the physics of large scales in multiple complementary ways.

Clusters of galaxies are defined as galaxy concentrations, but they can also 
be regarded as a specific class of objects in their own right. In the hierarchical framework of structure formation, they are the largest and last nonlinear objects in a quasi-equilibrium state. Observationally, they are well-defined and are systematically identified as strong X-ray emitters, thanks to their hot gas phase \citep[e.g.][]{Sarazin1998, Pierre2011}, and/or as isolated (red) galaxy overdensities \citep{Abell1958, Koester2007}, and/or via their imprint on the cosmological background radiation \citep{SZ1972, PlanckSZcatalogue, Reichardt2012}. Observed clusters are much more rare than individual galaxies (current cluster surveys amount up to a few tens of objects per deg$^2$), but the link to their parent dark matter halo mass is better understood. Most of cluster cosmological studies rely on the abundance of clusters and its evolution \citep{Borgani2001, Rozo2010, Mantz2010, Sehgal2011, Clerc2012}. However, very large ongoing surveys \citep[e.g.][]{Pierre2011} or planned surveys \citep{Predehl2009, Refregier2010} promise to overcome the statistical limit and better constrain cosmological models by stuyding the spatial clustering of clusters over large volumes.

This work follows our previous study \citep[][hereafter Paper~I]{Valageas2011} that analytically quantified covariance matrices for correlation functions in the framework of large galaxy cluster surveys. These calculations are essential to account for uncertainties in clustering analyses. Most generally, these uncertainties combine a shot-noise term (arising from the discreteness of the objects) and a sample variance term (due to the finite volume of the Universe being surveyed), as well as mixed contributions. All are related to the survey design and we applied our formalism to predict signal-to-noise ratios for observable quantities given different survey strategies \citep{Pierre2011, Valageas2011}.
We presented our method in a way as close as possible to observational analyses, e.g.~refering to the widely-used Landy-Szalay estimator for the two-point correlation function, integrated within fixed distance bins, averaging over all accessible redshift values, etc. Our results were found to agree well with large N-body simulations. However, only real-space comoving coordinates were considered, i.e., we ignored halo peculiar velocities. Because their radial component impacts object distance measurements via an additional spectral redshift, these peculiar velocities distort the correlation function estimates.

The goal of the present paper is to estimate the corrections induced by these redshift-space distortions, still considering galaxy clusters only, and to predict if these effects may be observed in present and future cluster surveys.
Indeed, redshift-space distortion measurements have attracted growing interest as a cosmological probe.
Schematically, this can be understood as follows: because peculiar velocities are linked to the gravitational well created by surrounding matter, bulk flows in the large scale structure are directly linked to the growth rate of structures. More precisely, redshift-space distortions are sensitive to $D_{+}(a)$, the linear growth factor, as a function of cosmic scale-factor $a$, and to $f=\dd\ln D_{+}/\dd \ln a$. In particular, observational results have been obtained by various galaxy spectroscopic surveys such as the 2dFGRS \citep{Peacock2001}, the SDSS \citep{Cabre2009,Reid2012}, 2dF-SDSS LRG and QSO survey \citep{Ross2007}, and the VVDS \cite{Guzzo2008}.

In the following, we assume galaxy clusters to have been detected individually in a dedicated survey (e.g.~in the X--ray band) and their redshift to have been subsequently determined, e.g.~by measurements of their member galaxy spectra and averaging over the corresponding galaxy redshifts. This is the most common -- but observationally expensive -- method to determine cluster redshifts. In practical applications, only a subsample of all cluster members can be observed due to observational constraints (limiting magnitude, compactness of the cluster in the instrumental focal plane, finite wavelength range of the spectrograph, identification of spectral lines, elimination of fore- and background objects, etc.). This translates into statistical errors on the cluster redshift estimate \citep[e.g.][]{Katgert1996, Adami2000}. Most generally, spectroscopic redshifts of galaxy clusters are considered to be precise to $\sigma(z) \sim 0.01 (1+z)$ or less.

To distinguish measurement errors from intrinsic covariances, we assume cluster redshifts to be known with infinite accuracy and neglect any kind of bias. 
Adding these sources of noise (which are not correlated with the shot-noise and sample-variance terms that we consider here) is not difficult as one only needs to add
the relevant covariance matrices. We do not consider this second step here, which
depends on the details of the instrument.

As in Paper~I, we considered in our calculations the effects of shot-noise and sample variance. We expanded the angular anisotropy of the redshift-space correlation function on Legendre polynomials. Our formalism then leads to expressions for the monopole and the two multipoles $2 \ell = 2$ and $2 \ell =4$. We provide explicit expressions for the expected quantities and their covariance matrices, taking into account the Kaiser effect and Gaussian as well as non-Gaussian contributions. Keeping in mind the application to galaxy cluster surveys, all quantities were integrated within the $0<z<1$ redshift range. Halos were defined by their mass $M$ and we describe the population by using a mass function and a bias model from previous works. The three- and four-point correlation function were estimated by means of a hierarchical model \citep{Peebles1980,Bernstein1994}, where those quantities only depend on combinations of two-point correlation functions. Comparisons with numerical simulations were made to assess the reliability of our analytical calculations.

Therefore, our main improvements over previous works consist in the inclusion of Gaussian and non-Gaussian contributions in the computation of covariance matrices, as well as in the derivation of observables close to those expected in practical studies focused on galaxy clusters.

This paper is organized as follows.
In Sect.~\ref{Halo-density} we first briefly describe the analytic models that
we used to estimate the means and covariance matrices of redshift-space
halo correlation functions, as well as the numerical simulations that we used to
check our results.
Then, focusing on the Landy \& Szalay estimator, we study the means of the 
various multipoles of the redshift-space correlation in
Sect.~\ref{Two-point-correlation}, and their covariance matrices in
Sect.~\ref{Covariance-matrix}. We compare our analytical results with
numerical simulations for the two lowest multipoles (the ``$2\ell=0$'' monopole
and the ``$2\ell=2$'' quadrupole).
We apply our formalism to several real surveys in Sect.~\ref{Applications},
for the three multipoles generated by the Kaiser effect
($2\ell=0,2,$ and $4$).
We conclude in Sect.~\ref{Conclusion}.

We give details of our calculations in Appendices~\ref{geometrical-averages}
and \ref{integrals}, and we show the auto- and cross-correlation matrices
of the ``$2\ell=2$'' multipole in Appendix~\ref{Correlation-matrix-l=2}.

\section{Redshift-space halo density field}
\label{Halo-density}

Before we describe our analysis of the covariance matrices for halo number
counts and correlation functions, we present in this section the analytic
models that we used for the underlying halo distributions (mass function and bias,
etc.) and the numerical simulations that we used to validate our results.

\subsection{Analytic models}
\label{Analytic-models}

To be consistent with the numerical simulations, we use the WMAP3 cosmology 
in Sects.~\ref{Two-point-correlation} and \ref{Covariance-matrix}
where we develop our formalism and compare our results with simulations,
that is, $\Omega_{\rm m}=0.24$, $\Omega_{\rm de}=0.76$,
$\Omega_b=0.042$, $h=0.73$, $\sigma_8=0.77$, $n_s=0.958$, and
$w_{\rm de}=-1$ \citep{Spergel2007}.
In Sect.~\ref{Applications}, where we apply our formalism to obtain forecasts
for current and future surveys, we use the more recent WMAP7 cosmology
\citep{Komatsu2011}, that is, $\Omega_{\rm m}=0.274$, 
$\Omega_{\rm de}=0.726$, $\Omega_b=0.046$, $h=0.702$, 
$\sigma_8=0.816$, $n_s=0.968$, and $w_{\rm de}=-1$.
 
Keeping in mind the study of X-ray clusters, we considered the
number counts and correlations of dark matter halos defined by the
nonlinear density contrast $\delta=200$. These halos are fully characterized
by their mass, and we did not investigate the relationship between this mass 
and cluster properties such as the gas temperature and X-ray luminosity.
These scaling laws can be added to our formalism to derive the cluster
number counts and correlations, depending on the quantities that are actually
measured, but we kept a more general setting in this paper.

\subsubsection{Halo mass function and bias}
\label{Halo-mass-function}

We use the halo mass function, $\dd n/\dd \ln M$, of \citet{Tinker2008}, 
and the halo bias, $b(M)$, of \citet{Tinker2010}. Then, we write the two-point
real-space correlation function between two halos labeled ``$i$'' and ``$j$'' as
\beq
\xih_{i,j} = b_i b_j \, \xi(|\vx_i-\vx_j|;z) ,
\label{xi-ij-bb}
\eeq
where $\xi$ is the real-space matter density correlation.
This corresponds to the halo power spectrum
\beq
P^{\rm h}_{i,j}(k;z) =  b_i b_j \, P(k;z) ,
\label{Pk-halo}
\eeq
where $P(k)$ is the matter density power spectrum.
Here we used the approximation of scale-independent halo bias, which is valid to
better than $10\%$ on scales $20<x<130 h^{-1}$ Mpc \citep{Manera2010},
with a small feature on the baryon acoustic scale ($r \sim 100 h^{-1}$Mpc)
of amplitude of $5\%$ \citep{Desjacques2010b}.

To simplify notations, we define the mean cumulative number density
$\nb$ of objects observed at a given redshift, within the mass bin $[M_-,M_+]$,
as
\beq
\nb(z) =  \int_{M_-}^{M_+} \frac{\dd M}{M} \, 
\frac{\dd n}{\dd\ln M}(M,z) ,
\label{Nbz-def}
\eeq
and the mean bias $\bb$ as
\beq
\bb(z) \, \nb(z) = \int_{M_-}^{M_+} \frac{\dd M}{M} \, b(M,z) \,
\frac{\dd n}{\dd\ln M}(M,z) .
\label{bb-def}
\eeq
In the following we omit to write the explicit boundaries on mass since we only
consider a single mass bin. It is straightforward to extend all our results to
the case of several mass bins by writing the relevant mass boundaries and
replacing factors such as $\bb^2\nb^2$ by $\bb_i \nb_i \bb_j \nb_j$.

\subsubsection{Redshift-space correlation functions}
\label{redshift-space}

We denote by $\vx$ and $\vs$ real-space and redshift-space coordinates.
They only differ by their radial component because of the peculiar velocity
along the line-of-sight,
\beq
\vs= \vx+  \frac{\ve_z\cdot\vv}{aH} \; \ve_z .
\label{s-x}
\eeq
Here, $\ve_z$ is the unit vector along the line of sight and $\vv$ the peculiar velocity.
At linear order, the peculiar velocity field and the density contrast are related in
Fourier space as
\beq
\tvv(\vk) = \ii \; a\frac{\dd\ln{D}_+}{\dd t} \; \frac{\vk}{k^2} \; \tdelta(\vk) ,
\label{vk}
\eeq
where $D_+$ is the linear growing mode. From the conservation of matter,
$(1+\delta^{(s)}) \dd\vs = (1+\delta) \dd\vx$, one obtains the redshift-space
matter power spectrum \citep{Kaiser1987}
\beq
P^{(s)}(\vk;z) = (1+f\mu_{\vk}^2)^2 \, P(k;z) , \;\;\; \mu_{\vk} = \frac{\ve_z\cdot\vk}{k} ,
\label{Ps-matter}
\eeq
where
\beq
f=\frac{\dd\ln D_+}{\dd\ln a} .
\label{f-def}
\eeq
Here and in the following, we use a superscript $(s)$ for redshift-space quantities.
Following common practice, we assume that the halo velocity field is not biased.
This yields at the same order the redshift-space halo power spectrum
\beq
P^{{\rm h}(s)}(\vk;z) = \bb^2 (1+\beta\mu_{\vk}^2)^2 \, P(k;z) ,
\label{Ps}
\eeq
with $\beta = f/\bb$, where $\bb$ is the mean bias of the halo population as
defined in Eq.(\ref{bb-def}) \citep{Percival2009}.
We write the associated redshift-space halo correlation function as
\beq
\xihs(\vs;z) = \bb^2 \, \xis(\vs;z) ,
\label{xis-ij-bb}
\eeq
where $\xis(\vs)$ can be expanded as \citep{Hamilton1992}
\beq
\xis(\vs) = \sum_{\ell=0}^2 \xi^{(s;2\ell)}(s) \, P_{2\ell}(\mu_{\vs})  ,
\;\;\; \mu_{\vs} = \frac{\ve_z\cdot\vs}{s} ,
\label{xis}
\eeq
where we introduced the functions
\beq
\xi^{(s;0)}(s) = \left(1+\frac{2\beta}{3}+\frac{\beta^2}{5}\right) \; \int_0^{\infty} \frac{\dd k}{k}
\Delta^2(k) \; j_0(k s) ,
\label{xis-0}
\eeq
\beq
\xi^{(s;2)}(s) = -\left(\frac{4\beta}{3}+\frac{4\beta^2}{7}\right) \; \int_0^{\infty} \frac{\dd k}{k}
\Delta^2(k) \; j_2(k s) ,
\label{xis-2}
\eeq
\beq
\xi^{(s;4)}(s) = \frac{8\beta^2}{35} \; \int_0^{\infty} \frac{\dd k}{k}
\Delta^2(k) \; j_4(k s) ,
\label{xis-4}
\eeq
where $j_{\ell}$ are the spherical Bessel functions and we introduced the
logarithmic power $\Delta^2(k)= 4\pi k^3 P(k)$.

The function $\xis$ and its multipoles $\xi^{(s;2\ell)}$ depend on the halo
population through the factor $\beta$ (i.e., the bias cannot be fully factored out as
in the real-space correlation (\ref{xi-ij-bb})).
However, it is convenient to introduce the auxiliary function $\xis$ as in
Eq.(\ref{xis-ij-bb}) so that most expressions obtained in the following have the
same form as in real space (which is recovered by setting $\beta=0$, that is,
$\xis=\xi$).

Independently of the linear-order approximation (\ref{Ps}), only even multipoles
can appear in the expansion (\ref{xis}) because of parity symmetry in
redshift-space. Nonlinear corrections may add higher-order even multipoles
that could be handled by a straightforward generalization of the approach
developed in this paper.

The redshift-space distortions (\ref{Ps}) and (\ref{xis}) only include the
``Kaiser effect'' \citep{Kaiser1987}, associated with large-scale flows (i.e., 
the collapse of overdense regions and the expansion of underdense regions), 
which amplifies the power because the density and velocity fields are closely 
related as in Eq.(\ref{vk}). However, on smaller scales within virialized regions, 
there appears a high velocity dispersion that adds a decorrelated random 
component to relation (\ref{s-x}). This yields the so-called
``fingers of God'' \citep{Jackson1972} that can be seen on galaxy surveys and
give a damping factor to their redshift-space power spectrum.
In this paper we focus on cluster surveys where this effect does not arise
(as can be checked in numerical simulations, \citep{Nishimichi2011}).
Indeed, because clusters are the largest nonlinear objects, cluster pairs are still in
the process of moving closer (or farther appart in fewer cases) and have not had
time to fall within a common potential well and develop a high velocity dispersion
through several orbital revolutions.
Therefore, we only need to consider the Kaiser effect as in Eqs.(\ref{Ps}) and
(\ref{xis}).
This provides a good agreement with numerical simulations for $s>20 h^{-1}$Mpc
(within $10\%$, except for the multipole $2\ell=4$, which shows stronger
deviations from
linear theory, \cite{Reid2011}, but we will see that this multipole is not very important
for cluster studies), and for $k < 0.1 h$Mpc$^{-1}$ in Fourier space
\citep{Percival2009,Nishimichi2011}.

We assume that these expressions still provide a good approximation up to the
weakly nonlinear scales that are relevant for studies of cluster correlations
($\ga 6 h^{-1}$Mpc). For numerical purposes we use the popular fitting 
formula to numerical simulations of \citet{Smith2003} for the matter power 
spectrum.

Nevertheless, the formalism that we develop in the following sections
is fully general and does not rely on Eqs.(\ref{xis-0})-(\ref{xis-4}) and the
truncation of the series (\ref{xis}) at $2\ell=4$. It can be applied to any
model for the redshift-space correlation $\xi^{(s)}(\vs)$. If this leads to
higher-order multipoles $\xi^{(s;2\ell)}$, one simply needs to extend the sums
to all required multipoles in the various expressions given in
Appendix~\ref{geometrical-averages}.

\subsubsection{Three-point and four-point halo correlations}
\label{three-point}

The covariance matrices of estimators of the halo two-point correlation
involve the three-point and four-point halo correlations. Therefore, we need
a model for these higher-order correlations.
As in \citet{Valageas2011}, following  \citet{Bernstein1994}, \citet{Szapudi1996}, 
and \citet{Meiksin1999}, we use a hierarchical ansatz 
\citep{Groth1977,Peebles1980}
and we write the real-space halo three-point correlation $\zetah$ as
\beq
\zetah_{1,2,3} = b_1 b_2 b_3 \; \frac{S_3}{3} \; \left[ \xi_{1,2} \xi_{1,3} 
+ \xi_{2,1} \xi_{2,3} + \xi_{3,1} \xi_{3,2} \right] ,
\label{zeta-def}
\eeq
where we sum over all three possible configurations over the three halos
labeled ``1'', ``2'', and ``3'', and the real-space halo four-point correlation
$\etah$ as
\beqa
\etah_{1,2,3,4} & = & b_1 b_2 b_3 b_4 \; \frac{S_4}{16} \;
\left[ \xi_{1,2} \xi_{1,3} \xi_{1,4} + 3 \, {\rm cyc.} \right. \nonumber \\
&& \left. + \xi_{1,2} \xi_{2,3} \xi_{3,4} + 11 \, {\rm cyc.} \right] ,
\label{eta-def}
\eeqa
where ``3 cyc.'' and ``11 cyc.'' stand for three and eleven terms that are
obtained from the previous one by permutations over the labels ``1,2,3,4''
of the four halos.
As in \citet{Valageas2011}, we take for the normalization factors $S_3$
and $S_4$ their large-scale limit, which is obtained by 
perturbation theory \citep{Peebles1980,Fry1984a,Bernardeau2002},
\beq
S_3 = \frac{34}{7} - (n+3) ,
\label{S3-def}
\eeq
\beq
S_4 = \frac{60712}{1323} - \frac{62}{3} (n+3) + \frac{7}{3} (n+3)^2 ,
\label{S4-def}
\eeq
where $n$ is the slope of the linear power spectrum at the scale of interest.
The model (\ref{zeta-def})-(\ref{S4-def}) is the simplest model that
qualitatively agrees with large-scale theoretical predictions
(which also give the scalings $\zeta \sim \xi^2$ and $\eta \sim \xi^3$,
but where the normalization factors $S_3$ and $S_4$ show an additional
angular dependence) and small-scale numerical results \citep{Colombi1996}
(which however show an additional moderate scale-dependence of the
normalization factors $S_3$ and $S_4$).
As shown in \citet{Valageas2011}, this model (\ref{zeta-def})-(\ref{S4-def})
provides reasonably good predictions for the covariance matrices of estimators
of the cluster two-point correlations (which focus on weakly nonlinear scales).

Next, for the redshift-space correlations $\zeta^{{\rm h}(s)}$ and
$\eta^{{\rm h}(s)}$ we use the same hierarchical ansatz
(\ref{zeta-def})-(\ref{eta-def}), where in the right-hand-side we replace $\xi$
by the redshift-space two-point correlation $\xis$ given in
Eq.(\ref{xis}).
This redshift-space model has not been checked in details against numerical 
simulations in previous works and we compare our predictions with
simulations in Sect.~\ref{Covariance-matrix} below.
We will check that it yields a reasonable agreement for covariance matrices
(especially along the diagonal) as for our purposes we only need
a simple and efficient approximation. 
If one aims at predicting the three-point and four-point redshift-space
correlations for their own sake, up to a high accuracy, one should
perform dedicated tests for these quantities and probably build more
sophisticated models, but this is beyond the scope of this paper.

This approximation allows us to extend in a straightforward fashion the
formalism developed in \citet{Valageas2011} to redshift-space estimators.
The only difference comes from the angular dependence of the two-point
correlation (\ref{xis}), that is, from the second and fourth spherical harmonics
$\xi^{(s;2)}(s) P_2(\mu)$ and $\xi^{(s;4)}(s) P_4(\mu)$.
In particular, as clearly seen in Eq.(\ref{Ps}), by setting $\beta=0$, which only
leaves the monopole contribution $\ell=0$, we recover the real-space results
given in \citet{Valageas2011}.

\subsection{Numerical simulations}
\label{simulations}

To check our analytical model, we compare our predictions with the numerical
simulations that we used in \citet{Valageas2011} for real-space statistics.
While we had considered only real-space coordinates in our previous study,
we used the peculiar velocity assigned to each halo
to conduct our comparison in redshift-space coordinates.

Here we briefly recall the main characteristics of these simulations.
We used the high-resolution, all-sky, ``horizon'' simulation run \citep{Teyssier2009}.
It consists of a $(2 h^{-1}\rm{Gpc})^3$ N-body simulation carried out
with the RAMSES code \citep{Teyssier2002}.
The effect of baryons is neglected and the total number of particles
in the simulation sets the total mass of the simulation.
The mass resolution, which is the mass of each of the $4096^3$ simulated particules,
is $\sim 10^{10}$\,$M_{\odot}$.
Halos are found using the HOP algorithm \citep{Eisenstein1998b}.
The (real-space) comoving position and peculiar velocity of each
halo are obtained by averaging over all particles making up its total mass.
Comoving coordinates are further transformed into sky positions
(two angles plus redshift) and redshift-space coordinates taking into account their radial velocity.
To ensure completeness of the simulation in all directions, we only considered
halos at $z<0.8$.

Simulated surveys are extracted from the full-sky light-cone catalog by defining non-overlapping 
rectangular regions in angular coordinates. Each field covers an area of 50 deg$^2$.
To minimize correlations between neighboring fields, we required a separation of 
20 deg between two consecutive surveys, which results in a total of 34 such independent 
fields distributed all across the sky.
Only for the results discussed in Section~\ref{Covariance-2l=2} we also used a 
10 deg gap (resulting in 138 fields).
For the purpose of deriving the correlation function, auxiliary random fields were created by 
shuffling the angular coordinates of halos in the original data fields. This operation 
preserves the original redshift and mass distribution of the simulation. The density of objects in the 
random fields was increased to one hundred 
times that of original data fields to avoid introducing spurious noise.
The Landy-Szalay 
estimator and its generalization (see Section.~\ref{Analytical-mean}, Eq.~\ref{hxis-DR-li}) provide an estimate of the two-point correlation and its multipoles in redshift space.
As in \citet{Valageas2011}, the mean and covariance of all derived quantities were finally estimated 
by sample averaging over the extracted fields.

\section{Mean redshift-space two-point correlation functions}
\label{Two-point-correlation}

Since we have in mind the application to cluster surveys and,
more generally, to deep surveys of rare objects, we considered
3D correlation functions averaged over a wide redshift bin (to
accumulate a large enough number of objects), rather than the usual
local 3D correlation functions at a given redshift.
This means that the quantities that we considered, while being
truly 3D correlations and not 2D angular correlations, nevertheless involve
integrations along the line of sight within a finite redshift interval.
We followed the formalism and the notations of \citet{Valageas2011} to derive the
means and the covariance matrices of estimators of these integrated redshift-space 
correlations.

\subsection{Analytical results}
\label{Analytical-mean}

We focus on the Landy \& Szalay estimator of the two-point correlation
\citet{Landy1993}, given by
\beq
\hxi = \frac{DD-2DR+RR}{RR} ,
\label{hxis-DR}
\eeq
where $D$ represents the data field and $R$ an independent Poisson
distribution, both with the same mean density.
In practice, before appropriate rescaling, the mean number density of the
Poisson process $R$ is actually taken to be much higher than the observed one,
so that the contribution from fluctuations of the denominator $RR$ to the noise of
$\hxi$ can be ignored.
The advantage of Eq.(\ref{hxis-DR}) is that one automatically includes
the geometry of the survey (including boundary effects, cuts, etc.),
because the auxiliary field $R$ is drawn on the same geometry.
In practice, our generalization of the estimator (\ref{hxis-DR}) to redshift-space
multipoles writes as
\beqa
\hxi^{(s)}_{R_i,2\ell_i} & = & \frac{4\ell_i\!+\!1}{\sum_{rr} 1} \biggl \lbrace \sum_{dd} 
P_{2\ell_i}(\mu_{dd})  - 2 \sum_{dr} P_{2\ell_i}(\mu_{dr}) \nonumber \\
&& + \sum_{rr} P_{2\ell_i}(\mu_{rr}) \biggl \rbrace ,
\label{hxis-DR-li}
\eeqa
where each sum counts all the pairs, $dd,dr$, or $rr$, within the samples $D$ and
$R$, that are separated by a redshift-space distance $s$ within the bin
$\Rim<s<\Rip$.
Compared with the usual monopole estimator (\ref{hxis-DR}), we added the 
geometrical weight $(4\ell_i\!+\!1)P_{2\ell_i}(\mu)$ to each pair
in the numerator while in the denominator we kept the unit weight.
Of course, for $\ell_i=0$ we recover Eq.(\ref{hxis-DR}).

Within our framework, we write this estimator $\hxi^{(s)}_{R_i,2\ell_i}$, for the multipole $2\ell_i$
of the mean correlation on scales delimited by $\Rim$ and $\Rip$, integrated over some redshift
range and mass interval, as
\beqa
\hxi^{(s)}_{R_i,2\ell_i} & \!\! = & \frac{1}{\QQ_i} \!\! \int \!\! \dd z \, 
\frac{\dd\chi}{\dd z}  \cD^2 \int \!\! \frac{\dd\vOm}{(\Delta\Omega)} 
\int \!\! \frac{\dd M}{M} \int_i \! \dd\vs'  (4\ell_i\!+\!1) P_{2\ell_i}(\mu')
\nonumber \\
&& \hspace{-1cm} \times \!\! \int \! \frac{\dd M'}{M'} \frac{\dd\hn}{\dd\ln M} 
\frac{\dd\hn}{\dd\ln M'} - \frac{2}{\QQ_i} \! \int \!\! \dd z \, \frac{\dd\chi}{\dd z}  \cD^2
\int \!\! \frac{\dd\vOm}{(\Delta\Omega)} \int \!\! \frac{\dd M}{M} \nonumber \\
&& \hspace{-1cm} \times \!\! \int_i \! \dd\vs' (4\ell_i\!+\!1) P_{2\ell_i}(\mu') \!
\int \!\! \frac{\dd M'}{M'} \,  \frac{\dd\hn}{\dd\ln M} \frac{\dd n}{\dd\ln M'} 
+ \delta_{\ell_i,0} ,
\label{xis-1}
\eeqa
with
\beqa
\QQ_i & = & \int\dd z \, \frac{\dd\chi}{\dd z} \, \cD^2
\int\frac{\dd\vOm}{(\Delta\Omega)} \int\frac{\dd M}{M} 
\int_i \dd\vs' \int\frac{\dd M'}{M'}  \nonumber \\
&& \times \, \frac{\dd n}{\dd\ln M} \frac{\dd n}{\dd\ln M'} .
\label{QQi-def}
\eeqa
Here we denoted $\int_i \dd\vs'$ as the integral $\int_{\Rim}^{\Rip}\dd\vs'$ over
the 3D spherical shell $i$, of inner and outer radii $\Rim$ and $\Rip$ in
redshift-space, $\mu'=(\ve_z\cdot\vs')/s'$ is the cosine of the angle between
the line of sight and the halo pair,
$\chi(z)$ and $\cD(z)$ are the comoving radial and angular distances,
$\delta_{\ell_i,0}$ is the Kronecker symbol (which is unity if $\ell_i=0$ and zero
otherwise),
and $\dd\hn/\dd\ln M$ is the observed density of objects.
Here and in the following, we note observed
quantities by a hat (i.e., in one realization of the sky) to distinguish them from mean
quantities, such as the mean comoving number density $\dd n/\dd\ln M$.
The two subscripts $R_i$ and $2\ell_i$ label the radial bin and the multipole.
If we consider $N_R$ radial bins and $N_{\ell}$ multipoles, there are
$N_R\times N_{\ell}$ values of the index $i$ (i.e., two different indices $i$ and
$j$ may correspond to different radial bins and multipoles, or to the same radial
bin but with two different multipoles, and vice versa).

In Eqs.(\ref{xis-1}) and (\ref{QQi-def}) we considered a survey with a single 
window of angular area $(\Delta\Omega)$ and a single mass bin.
It is straigtforward to generalize to several distant (uncorrelated) angular windows
and to several mass bins.
The redshift interval $\Delta z$ is not necessarily small, and to increase the
statistics we can choose the whole redshift range of the survey.
It is again straigtforward to generalize to several nonoverlapping redshift intervals,
if they are large enough to neglect cross-correlations between different bins 
($\Delta z \ga 0.1$, see Fig.~11 in \citet{Valageas2011}).

As in Eq.(\ref{hxis-DR-li}), the counting method that underlies the first term in
Eq.(\ref{xis-1}) can be understood as follows.
We span all objects in the ``volume'' $(z,\vOm,\ln M)$, and count all
neighbors at distance $s'$, within the shell $[\Rim,\Rip]$, with a mass
$M'$. We denote with unprimed letters the quantities associated with the
first object, $(z,\vOm,\ln M)$, and with primed letters the quantities associated
with the neighbor of mass $M'$ at distance $s'$.
Thus, with obvious notations, $\dd\hn/\dd\ln M$ and $\dd\hn/\dd\ln M'$
are the observed number densities at the first and second (neighboring) points.
In contrast, the denominator $\QQ_i$ involves the mean number densities
$\dd n/\dd\ln M$ and $\dd n/\dd\ln M'$. Therefore, $\QQ_i$ is not a random quantity
and accordingly it shows no noise. 
Similarly, the difference between the terms associated with $DD$ and $DR$ is that
in the former we have a product of two observed number densities,
$(\dd\hn/\dd\ln M)\times(\dd\hn/\dd\ln M')$, while in the latter we have
a crossproduct between the observed and the mean number densities,
$(\dd\hn/\dd\ln M)\times(\dd n/\dd\ln M')$.

Then, if $\ell_i \neq 0$ the second and third terms in Eq.(\ref{xis-1}) vanish. 
This also means that for $\ell_i \neq 0$ the Landy \& Szalay estimator 
(\ref{hxis-DR-li}) and the Peebles \& Hauser estimator \citep{Peebles1974a},
which would read as $\hxi=DD/RR-\delta_{\ell_i,0}$ (with the implicit
factor $(4\ell_i\!+\!1)P_{2\ell_i}$ in $DD$), are equivalent (in the limit where the
density of the auxiliary random sample $R$ goes to infinity).

The angular factor $(4\ell_i+1) P_{2\ell_i}(\mu')$ ensures that the estimator
(\ref{xis-1}) extracts the multipole $2\ell_i$ of the redshift-space correlation,
as in (\ref{xis}).
In our previous analysis of the real-space correlation \citep{Valageas2011} 
we only needed to consider the monopole term, $\ell_i=0$, because higher multipoles
vanished. Here, because the redshift-space correlation is anisotropic, higher 
multipoles are nonzero and contain valuable information. As seen in
Eqs.(\ref{xis-0})-(\ref{xis-4}), they constrain the parameter $\beta=f/\bb$, and
in turn the growth rate of the density fluctuations and the bias of the halos.

The quantity $\QQ_i$ also writes as\footnote{We can neglect the impact of the
change from $\vx$ to $\vs$ on background cosmological quantities such
as the radial and angular distances $\chi$ and $\cD$, which vary on scales on the
order of $c/H_0$. Redshift-space distortions only have a significant effect on the
correlation $\xis$ of the halo population, which is studied on smaller scales
$\sim 20 h^{-1}$ Mpc and becomes anisotropic.} 
\beq
\QQ_i = \int \dd\chi \, \cD^2 \, \nb^2 \, \cV_i ,
\label{QQ-1}
\eeq
where the volume $\cV_i$ of the $i$-shell is
\beq
\cV_i(z) = \frac{4\pi}{3} [\Rip(z)^3-\Rim(z)^3] ,
\label{Vi-def}
\eeq
which may depend on $z$. In practice, one would usually choose constant
comoving shells, so that $\cV_i$ does not depend on $z$.
To obtain Eq.(\ref{QQ-1}) we used that $\dd n/\dd\ln M$ and
$\dd n/\dd\ln M'$ have no scale dependence (because they correspond to a
uniform distribution of objects) and we neglected edge
effects. (These finite-size effects are discussed and evaluated in
Appendix~B of \citet{Valageas2011}).
Next, the mean of the estimator $\hxi^{(s)}$ reads as 
\beq
\lag\hxi^{(s)}_{R_i,2\ell_i}\rag = \frac{1}{Q_i} \int\dd\chi \, \cD^2 \, 
\bb^2 \, \nb^2 \, \cV_i \, \overline{\xis_{R_{i},2\ell_i}}(z) ,
\label{xi-3}
\eeq
with
\beq
\overline{\xis_{R_{i},2\ell_i}}(z) = \int_i\frac{\dd\vs'}{\cV_i} \, (4\ell_i+1) 
P_{2\ell_i}(\mu') \, \xis(\vs';z) ,
\label{xi-i-i'-def}
\eeq
which is the radial average of the $2\ell_i$-multipole of $\xi^{(s)}$, over the 3D 
spherical shell associated with the radial bin $i$ in redshift space.
Indeed, from the multipole expansion (\ref{xis}) this also reads as
\beq
\overline{\xis_{R_{i},2\ell_i}} = \int_{\Rim}^{\Rip} \frac{\dd r \; 3 r^2}{\Rip^3-\Rim^3}
\;\xi^{(s;2\ell_i)}(r) .
\label{xi-i-i'-1}
\eeq

\subsection{Comparison with simulations}
\label{Comparison-mean}

\begin{figure}
\begin{center}
\epsfxsize=8.5 cm \epsfysize=6.5 cm {\epsfbox{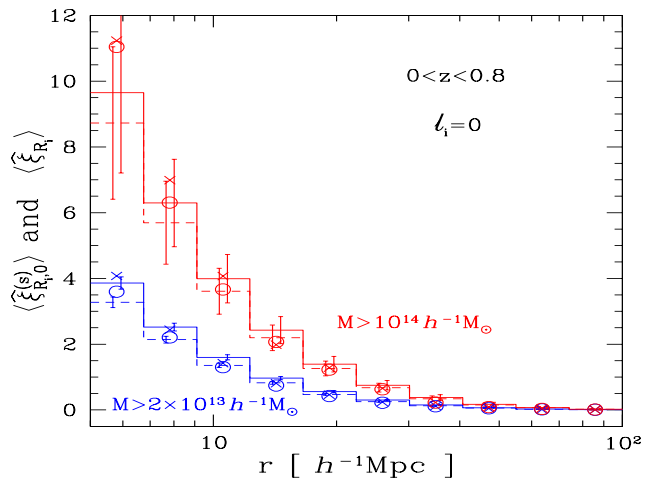}}
\end{center}
\caption{Mean redshift-space (solid lines) and real-space (dashed lines)
halo correlations, $\lag\hxi^{(s)}_{R_i,0}\rag$ and $\lag\hxi_{R_i}\rag$.
We considered ten comoving distance bins, within $5<r<100 h^{-1}$Mpc, 
equally spaced in $\log(r)$. We counted all halos above the thresholds 
$M_*=2\times 10^{13}$ and $10^{14} h^{-1} M_{\odot}$, 
within the redshift interval $0<z<0.8$.
We compare our analytical results with numerical simulations
(crosses for the monopole redshift-space correlation and circles for the
real-space correlation).
The $3-\sigma$ error bars (slightly shifted to the left and to the right for the
real-space and redshift-space correlations) are obtained from the analytical 
covariance matrices derived in Sect.~\ref{Covariance-matrix}.}
\label{fig_XisR}
\end{figure}

\begin{figure}
\begin{center}
\epsfxsize=8.5 cm \epsfysize=6.5 cm {\epsfbox{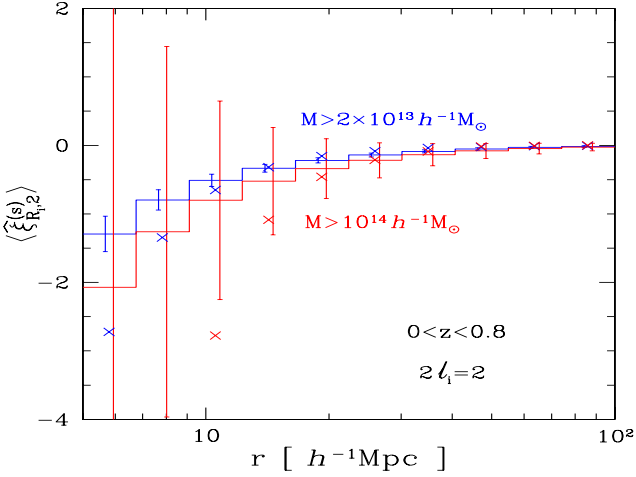}}
\end{center}
\caption{``$2\ell\!=\!2$'' redshift-space halo correlation
$\lag\hxi^{(s)}_{R_i,2}\rag$. As in Fig.~\ref{fig_XisR},
we considered ten comoving distance bins, within $5<r<100 h^{-1}$Mpc, 
equally spaced in $\log(r)$, and counted all halos above the thresholds 
$M_*=2\times 10^{13}$ and $10^{14} h^{-1} M_{\odot}$, 
within the redshift interval $0<z<0.8$.
We compare our analytical results with numerical simulations (crosses).
The $3-\sigma$ error bars (slightly shifted to the left and to the right for the
real-space and redshift-space correlations) are obtained from the analytical 
covariance matrices derived in Sect.~\ref{Covariance-matrix}.}
\label{fig_XisR_l1l1}
\end{figure}

We show in Fig.~\ref{fig_XisR} the mean redshift-space and real-space 
halo correlations, $\lag\hxi^{(s)}_{R_i,0}\rag$ and $\lag\hxi_{R_i}\rag$,
over ten distance bins. We find that because of the prefactor
$(1+2\beta/3+\beta^2/5)$ in Eq.(\ref{xis-0})
the monopole redshift-space correlation is greater than the
real-space correlation by
$10\%$ for massive halos, $M>10^{14} h^{-1} M_{\odot}$, and by
$18\%$ for $M>2\times 10^{13} h^{-1} M_{\odot}$,
in the redshift range $0<z<0.8$.
The relative redshift-space distortion is weaker for more massive halos 
because they have a larger bias $\bb$. This decreases the factor
$\beta=f/\bb$.

Although the results from numerical simulations are somewhat noisy, they
agree reasonably well with our analytical predictions. They give the same
order of magnitude as our predictions for both the redshift-space and
real-space correlations. In particular, they fall within the expected 
$3-\sigma$ error-bars from our analytical predictions (these error bars
are obtained from the analytical covariance matrices derived in 
Sect.~\ref{Covariance-matrix} below).
This also shows that the accurate computation of such quantities remains
a difficult task for numerical simulations (in this case it would require many 
realizations of the simulated sky resulting in a heavy computational load)
and it is useful to have analytical models for cross-checking.
In particular, while the ratio $\lag\hxi^{(s)}_{R_i,0}\rag/\lag\hxi_{R_i}\rag$
is greater than unity and smooth within the analytical model (it is actually
scale-independent in our simple model (\ref{xis-0})), it shows spurious
fluctuations in the simulation data for rare massive halos because of low
statistics.

For practical purposes, Fig.~\ref{fig_XisR} shows that
to constrain cosmology with the clustering of X-ray clusters it is
necessary to take into account redshift-space distortions if one aims at an
accuracy better than $10\%$.

We show in Fig.~\ref{fig_XisR_l1l1} our results for the ``$2\ell\!=\!2$''
redshift-space correlation $\xi^{(s)}_{R_i,2}$. 
This corresponds to the second term of
the expansion (\ref{xis}) over Legendre polynomials.
As seen from Eq.(\ref{xis-2}), this term is negative and its amplitude
is larger for more massive halos because of the factor 
$\bb^2 (4\beta/3+4\beta^2/7)= 4 f \bb/3+4 f^2/7$, which grows with the
halo bias.
However, because the error bars are much larger for rare massive halos,
the associated signal is more difficult to measure.
We obtain a good agreement with the simulations on large scales but
on small scales below $10 h^{-1}$Mpc our model seems to underestimate
the amplitude of this quadrupole component of the redshift-space correlation.
This may be partly due to higher-order contritbutions to Eq.(\ref{Ps-matter})
that modify the relationship between real-space and redshift-space power
spectra. For instance, within perturbation theory next-to-leading terms
at order $P_L^2$ arise and have been found to increase the amplitude of
the redshift-space power and to improve the agreement with simulations
\citep{Taruya2010,Nishimichi2011}.
However, we do not investigate such modifications in this paper.

We do not plot the last ``$2\ell=4$'' term, $\xi^{(s)}_{R_i,4}$, because
our analytical computations show that it is very small (smaller than $0.1$
on the scales shown in Figs.~\ref{fig_XisR} and \ref{fig_XisR_l1l1})
and within the error bars.
However, we have checked that our model agrees with the simulations for
$r>10 h^{-1}$ Mpc (on smaller scales our model again seems
to underestimate this correlation).

\section{Covariance matrices}
\label{Covariance-matrix}

\subsection{Explicit expressions}
\label{expressions}

We now consider the covariance matrices $C^{(s)}_{i,j}$ of the estimators
$\hxi^{(s)}_{R_i,2\ell_i}$, defined by
\beq
C^{(s)}_{i,j} =  \lag\hxi^{(s)}_{R_i,2\ell_i}\hxi^{(s)}_{R_j,2\ell_j}\rag - 
\lag\hxi^{(s)}_{R_i,2\ell_i}\rag \lag\hxi^{(s)}_{R_j,2\ell_j}\rag .
\label{Cij-def}
\eeq
To simplify notations, we do not write the labels $R_i$ and $2\ell_i$ in the
covariance matrices and the subscript $i$ again refers to both the radial bin,
$[\Rim,\Rip]$, and the multipole, $2\ell_i$.

The results of \citet{Valageas2011} directly extend to redshift space
and we obtain the expression
\beq
C^{(s)}_{i,j} = C^{(s)(2)}_{i,j} + C^{(s)(3)}_{i,j} + C^{(s)(4)}_{i,j} ,
\eeq
with (see also \citet{Landy1993,Szapudi2001a,Bernardeau2002a})
\beqa
C_{i,j}^{(s)(2)} & = & \delta_{R_i,R_j} \,
\frac{2(4\ell_i\!+\!1)(4\ell_j\!+\!1)}{(\Delta\Omega) \QQ_i^2} \int \dd\chi_i
\, \cD_i^2 \frac{\dd\vOm_i}{(\Delta\Omega)} \frac{\dd M_i}{M_i} 
\nonumber \\
&& \times  \int_i \dd\vs_{i'} \frac{\dd M_{i'}}{M_{i'}} \, \frac{\dd n}{\dd\ln M_i}
\frac{\dd n}{\dd\ln M_{i'}} \nonumber \\
&& \times P_{2\ell_i}(\mu_{i'}) P_{2\ell_j}(\mu_{i'}) \left[ 1+\xi^{{\rm h}(s)}_{i,i'} \right] ,
\label{C2-def}
\eeqa
\beqa
C_{i,j}^{(s)(3)} & \!\! = \! & \frac{4(4\ell_i\!+\!1)(4\ell_j\!+\!1)}
{(\Delta\Omega)\QQ_i\QQ_j} \! \int \!\! \dd\chi_i \, \cD_i^2 \frac{\dd\vOm_i}
{(\Delta\Omega)} \frac{\dd M_i}{M_i} \int_i \! \dd\vs_{i'} 
\frac{\dd M_{i'}}{M_{i'}} \nonumber \\
&& \times \int_j \dd\vs_{j'} \frac{\dd M_{j'}}{M_{j'}} \; \frac{\dd n}{\dd\ln M_i}
\frac{\dd n}{\dd\ln M_{i'}} \frac{\dd n}{\dd\ln M_{j'}} \nonumber \\
&& \times P_{2\ell_i}(\mu_{i'}) P_{2\ell_j}(\mu_{j'})
\left[ \xi^{{\rm h}(s)}_{i',j'} + \zeta^{{\rm h}(s)}_{i,i',j'} \right] ,
\label{C3-def}
\eeqa
\beqa
\!C_{i,j}^{(s)(4)} & \!\! = \! & \frac{(4\ell_i\!+\!1)(4\ell_j\!+\!1)}{\QQ_i\QQ_j} \! \int \!
\dd\chi_i \cD_i^2 \frac{\dd\vOm_i}{(\Delta\Omega)} \frac{\dd M_i}{M_i} 
\int_i \! \dd\vs_{i'} \frac{\dd M_{i'}}{M_{i'}} \nonumber \\
&& \hspace{-0.5cm} \times \frac{\dd n}{\dd\ln M_i} \frac{\dd n}{\dd\ln M_{i'}}
\int \! \dd\chi_j \cD_j^2 \frac{\dd\vOm_j}{(\Delta\Omega)} \frac{\dd M_j}{M_j}
\int_j \! \dd\vs_{j'} \frac{\dd M_{j'}}{M_{j'}} \nonumber \\
&&  \hspace{-0.5cm} \times \frac{\dd n}{\dd\ln M_j} \frac{\dd n}{\dd\ln M_{j'}} 
P_{2\ell_i}(\mu_{i'}) P_{2\ell_j}(\mu_{j'}) \nonumber \\
&&  \hspace{-0.5cm} \times \left[ 2 \xi^{{\rm h}(s)}_{i;j'} \xi^{{\rm h}(s)}_{i';j}
+ \eta^{{\rm h}(s)}_{i,i';j,j'} \right] .
\label{C4-def}
\eeqa
Here we used the same notation as in \citet{Valageas2011} that we recalled below 
Eq.(\ref{QQi-def}). Thus, labels $i$ and $j$ refer to objects that are at the center
of the $\cV_i$ and $\cV_j$ shells, whereas labels $i'$ and $j'$ refer to objects that
are within the $\cV_i$ and $\cV_j$ shells. The pairs $\{i,i'\}$ and $\{j,j'\}$
are located at unrelated positions in the observational cone, which we integrate
over.

In Eq.(\ref{C2-def}) we note $\delta_{R_i,R_j}$ the Kronecker symbol with
respect to the radial bins. Thus, $\delta_{R_i,R_j}$ is unity if
$[\Rim,\Rip]\equiv[\Rjm,\Rjp]$ and zero otherwise (we consider non-overlapping
radial bins), independently of the multipoles $\ell_i$ and $\ell_j$ (which may be
equal or different). This shot-noise contribution arises from cases where
the pairs $\{i,i'\}$ and $\{j,j'\}$ are the same (i.e., they involve the same two halos),
which implies the same pair length.

The label $C^{(n)}$ refers to quantities that involve $n$ distinct objects.
Thus, the contributions $C^{(2)}$
and $C^{(3)}$ arise from shot-noise effects (as is apparent through the prefactors
$1/(\Delta\Omega)$), associated with the discreteness
of the number density distribution, and they would vanish for continuous
distributions. However, they also involve the two-point
and three-point correlations, and as such they couple discreteness effects
with the underlying large-scale correlations of the population.
In case of zero large-scale correlations, $C^{(2)}$ remains nonzero because of the
unit factor in the brackets and becomes a purely shot-noise contribution,
arising solely from discreteness effects.
The contribution $C^{(4)}$ is a pure sample-variance contribution and
does not depend on the discreteness of the number density distribution
(hence there is no $1/(\Delta\Omega)$ prefactor).

We can reorganize these various terms into three contributions,
\beq
C^{(s)}_{i,j} = C^{(s){\rm (G)}}_{i,j} + C^{(s)(\zeta)}_{i,j} + C^{(s)(\eta)}_{i,j} ,
\label{Cij-G-NG}
\eeq
where $C^{(s){\rm (G)}}_{i,j}$ gathers all terms that only depend on the
two-point correlation, $C^{(s)(\zeta)}_{i,j}$ arises from the three-point correlation
(i.e., the last term in Eq.(\ref{C3-def})), and $C^{(s)(\eta)}_{i,j}$ arises from the 
four-point correlation (i.e., the last term in Eq.(\ref{C4-def})).
We obtain for the Gaussian contribution
\beqa
C_{i,j}^{(s){\rm (G)}} & = & \delta_{R_i,R_j} \frac{2}{(\Delta\Omega)\QQ_i} 
\left[ \delta_{\ell_i,\ell_j} (4\ell_i+1) + \lag\hxi^{(s)}_{R_i;2\ell_i,2\ell_j}\rag \right] 
\nonumber \\
&& + \frac{4}{(\Delta\Omega)\QQ_i\QQ_j} \int \dd\chi \, \cD^2 \, \bb^2 \, \nb^3 \, 
\cV_i \cV_j \, \overline{\xis_{i',j'}}  \nonumber \\
&& + \frac{2}{\QQ_i\QQ_j} \int \dd\chi \, \cD^5 \, \bb^4 \, \nb^4 \,
\cV_i \cV_j \, \overline{\xis_{i;j'} \xis_{i';j}} \;\; ,
\label{Cij-G}
\eeqa
which involves terms that are constant, linear, and quadratic over $\xis$.
Here we introduced the mean
\beq
\lag\hxi^{(s)}_{R_i;2\ell_i,2\ell_j}\rag =  \frac{1}{Q_i} \int\dd\chi \, \cD^2 \, 
\bb^2 \, \nb^2 \, \cV_i \, \overline{\xis_{R_{i};2\ell_i,2\ell_j}}(z) ,
\label{xi-Ri-li-lj}
\eeq
with
\beqa
\overline{\xis_{R_{i};2\ell_i,2\ell_j}}(z) & = & \int_i\frac{\dd\vs'}{\cV_i} \,
(4\ell_i\!+\!1) (4\ell_j\!+\!1) P_{2\ell_i}(\mu') P_{2\ell_j}(\mu') \nonumber \\
&& \times \; \xis(\vs';z) .
\label{xi-Ri-li-lj-def}
\eeqa
From the multipole expansion (\ref{xis}) this also reads as
\beqa
\overline{\xis_{R_{i};2\ell_i,2\ell_j}} & = & (4\ell_i\!+\!1) (4\ell_j\!+\!1)
\int_{\Rim}^{\Rip} \frac{\dd r \; 3 r^2}{\Rip^3-\Rim^3} \nonumber \\
&& \times \sum_{\ell=0}^2 
\left( \bea{ccc} 2\ell & 2\ell_i & 2\ell_j  \\ 0 & 0 & 0 \ea \right)^{\!\!2} \xi^{(s;2\ell)}(r) ,
\label{xi-Ri-li-lj-l}
\eeqa
which involves the square of the Wigner 3-j symbol.

Using the hierarchical ansatz described in Sect.~\ref{three-point}, the
contribution $C^{(s)(\zeta)}_{i,j}$ associated with the three-point correlation
writes as
\beqa
C_{i,j}^{(s)(\zeta)} & = & \frac{4}{(\Delta\Omega)\QQ_i\QQ_j} \int \dd\chi \,
\cD^2 \, \bb^3 \, \nb^3 \, \cV_i \cV_j \, \frac{S_3}{3}  \nonumber \\
&& \times \, \left[ \overline{\xis_{i'}} \times \overline{\xis_{j'}} 
+ \overline{\xis_{i',i} \xis_{i',j'}} +\overline{\xis_{j',i} \xis_{j',i'}} \right] ,
\label{Cij-zeta}
\eeqa
while the contribution $C^{(s)(\eta)}_{i,j}$ associated with the four-point correlation
writes as
\beqa
C_{i,j}^{(s)(\eta)} & = & \frac{2}{\QQ_i\QQ_j} \int \dd\chi \,
\cD^5 \, \bb^4 \, \nb^4 \, \cV_i \cV_j \, \frac{S_4}{16} \left[ \overline{\xis_{i'}}
\times \overline{\xi_{i;j} \xis_{i;j'}} \right . \nonumber \\
&& + \overline{\xis_{j'}} \times \overline{\xi_{i;j} \xis_{j;i'}} + 2 \,
\overline{\xis_{i'}} \times \overline{\xis_{j'}} \times \xicyl \nonumber \\
&& \left. + 2 \, \overline{\xis_{j';i}\xi_{i;j}\xis_{j;i'}} 
+ \overline{\xis_{j';i}\xis_{i,i'}\xis_{i';j}} + \overline{\xis_{i';j}\xis_{j,j'}\xis_{j';i}} 
 \right] .
\label{Cij-eta}
\eeqa
Here the quantities that are overlined are various geometrical averages of the
redshift-space two-point correlation functions. For instance, the geometrical
average introduced in the first integral in Eq.(\ref{Cij-G}) writes as
\beqa
\overline{\xis_{i',j'}}(z) & \! = \! & (4\ell_i\!+\!1) (4\ell_j\!+\!1) \int_i\frac{\dd\vs_{i'}}{\cV_i} 
P_{2\ell_i}(\mu_{i'}) \int_j \frac{\dd\vs_{j'}}{\cV_j} P_{2\ell_j}(\mu_{j'}) \nonumber \\
&& \times \; \xis(\vs_{i'}-\vs_{j'};z) .
\label{I3-ij-xi-def}
\eeqa
We give the explicit expressions of these various terms in
Appendix~\ref{geometrical-averages}, and we describe how angular integrations
can be performed using the spherical harmonic decomposition (\ref{xis}),
as well as the flat-sky and Limber's approximations.
All terms involve the redshift-space correlation $\xis$, except for $\xicyl$,
which only depends on the power along  the transverse directions (and is not
affected by redshift-space distortions).

\subsection{Covariance matrix of the monopole redshift-space correlation}
\label{Covariance-monopole}

In this section we study the covariance matrix $C^{(s)}_{i,j}$ of the
monopole redshift-space correlation, $\xi^{(s)}_{R_i,0}$, that is,
$\ell_i=\ell_j=0$.

\subsubsection{Comparison between redshift-space and real-space covariance
matrices}
\label{comparison}

\begin{figure}
\begin{center}
\epsfxsize=8.5 cm \epsfysize=6.5 cm {\epsfbox{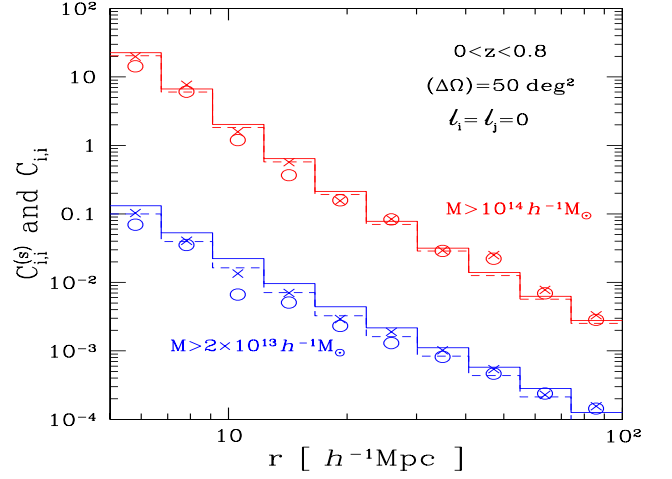}}
\end{center}
\caption{Redshift-space (solid line, for $2\ell=0$) and real-space 
(dashed line) covariance matrices, $C^{(s)}_{i,j}$ and $C_{i,j}$, along the 
diagonal, $i=j$. We show the results obtained for halos in the redshift range
$0<z<0.8$ with an angular window of $50$ deg$^2$.}
\label{fig_Cii}
\end{figure}

\begin{figure}
\begin{center}
\epsfxsize=8.5 cm \epsfysize=6.5 cm {\epsfbox{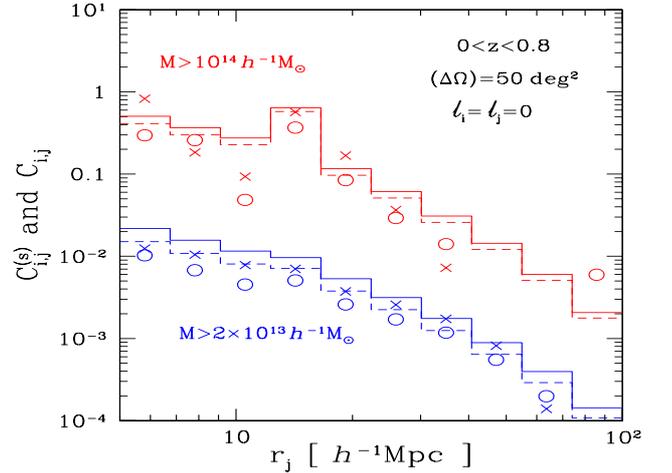}}
\end{center}
\caption{Redshift-space (solid line, for $2\ell=0$) and real-space (dashed 
line) covariance matrices, $C^{(s)}_{i,j}$ and $C_{i,j}$, as a function of $j$,
for $i=4$ associated with the distance bin $12.3<r<16.6h^{-1}$Mpc.
We show the results obtained for halos in the redshift range
$0<z<0.8$ with an angular window of $50$ deg$^2$.}
\label{fig_Cij}
\end{figure}

We show in Figs.~\ref{fig_Cii} and \ref{fig_Cij} the redshift-space and real-space
covariance matrices $C^{(s)}_{i,j}$ and $C_{i,j}$, for halos in the redshift range 
$0<z<0.8$.
The redshift-space covariance matrix is greater than the
real-space covariance matrix by about $11\%$ along the diagonal and by $20\%$
for off-diagonal entries, for massive halos,
$M>10^{14} h^{-1}M_{\odot}$.
For $M>2\times 10^{13} h^{-1}M_{\odot}$, these amplification factors
are higher, $33\%$ along the diagonal and $40\%$
for off-diagonal entries.
This could be expected from the higher value of the redshift-space correlation
function noticed in Fig.~\ref{fig_XisR}. This is due to the amplification of the
redshift-space power spectrum through the Kaiser effect, as shown by
Eq.(\ref{Ps}).
Again, the relative effect is weaker for rare massive halos because of their larger
bias $\bb$, which decreases the associated factor $\beta=f/\bb$.
The amplification of the covariance matrix is lower along the diagonal, where the
covariance matrix includes a pure shot-noise term (first term in Eq.(\ref{Cij-G}))
that is independent of $\xis$ and is not amplified by the Kaiser effect.

We obtain a reasonably good agreement with the numerical simulations for
both redshift-space and real-space covariance matrices, for diagonal and
off-diagonal entries. The simulation data are somewhat noisier for off
diagonal entries and show spurious oscillations in Fig.~\ref{fig_Cij}
that are not physical but due to low statistics, especially for rare massive
halos.
This shows that analytic models are competitive tools to estimate these
covariance matrices. In particular, for data analysis purposes it can be helpful
to have smooth covariance matrices, which make matrix inversion more
reliable.

In any case, redshift-space distortions only have a moderate impact on the
covariance matrices and do not change their magnitude.
This means that most of the real-space results obtained in \citet{Valageas2011}
remain valid.

\begin{figure}
\begin{center}
\epsfxsize=8.5 cm \epsfysize=6.5 cm {\epsfbox{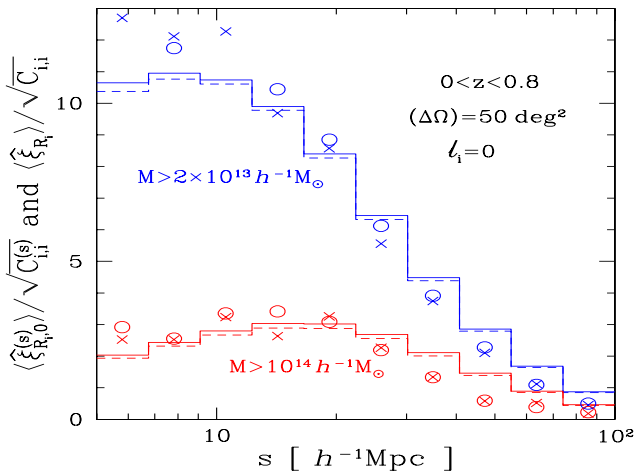}}
\end{center}
\caption{Signal-to-noise ratio of the redshift-space (solid line) and
real-space (dashed line) two-point correlations.
We show the results obtained for halos in the redshift range
$0<z<0.8$ with an angular window of $50$ deg$^2$.}
\label{fig_SN}
\end{figure}

We show in Fig.~\ref{fig_SN} an estimate of the signal-to-noise ratio of the
redshift-space and real-space two-point correlations, defined as
$\lag\hxi_{R_i}\rag/\sqrt{C_{i,i}}$.
A more precise analysis would require using the full covariance matrix, but
this ratio should allow us to check whether redshift distortions, which amplify
both the mean correlation and its covariance matrix, degrade the
signal-to-noise ratio. Clearly, Fig.~\ref{fig_SN} shows that the 
signal-to-noise ratio is not strongly affected by the Kaiser effect. Indeed, the
amplification of $\hxi^{(s)}_{R_i,0}$ overweights the amplification of
$C^{(s)}_{i,j}$ and the signal-to-noise ratio slightly increases.
This can be understood because the contributions to $C^{(s)}_{i,j}$
that are cubic over $\xis$ (or of higher order), which in our framework
only come from the four-point correlation contribution (\ref{Cij-eta}) to the
pure sample-variance term, are less important than the mixed shot-noise
and sample-variance contributions that are constant or linear over $\xis$.

The results found in Fig.~\ref{fig_SN}  mean that redshift distortions do not 
prevent measuring the clustering of X-ray clusters and its use to constrain 
cosmology, since the signal-to-noise ratio remains as high and even slightly 
higher.

\subsubsection{Importance of non-Gaussian contributions}
\label{non-Gaussian}

\begin{figure}
\begin{center}
\epsfxsize=8.5 cm \epsfysize=6.5 cm {\epsfbox{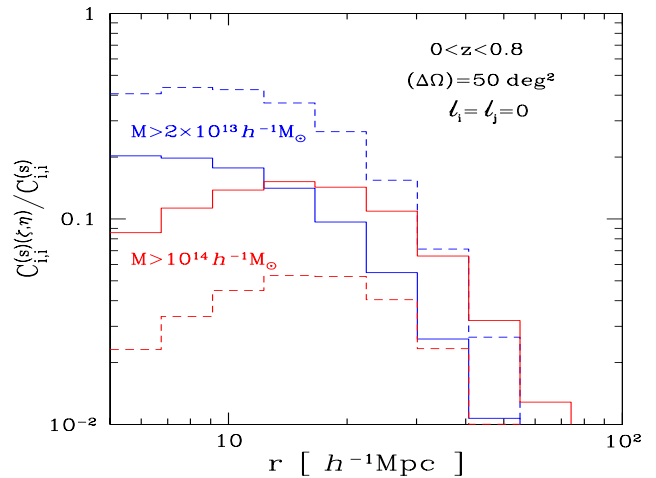}}
\end{center}
\caption{Relative importance of the non-Gaussian contributions
$C^{(s)(\zeta)}_{i,j}$ (solid line) and $C^{(s)(\eta)}_{i,j}$ (dashed line)
to the redshift-space covariance matrix $C^{(s)}_{i,j}$, along the diagonal, $i=j$.
We show the results obtained for halos in the redshift range
$0<z<0.8$ with an angular window of $50$ deg$^2$.}
\label{fig_Cii_NG}
\end{figure}

\begin{figure}
\begin{center}
\epsfxsize=8.5 cm \epsfysize=6.5 cm {\epsfbox{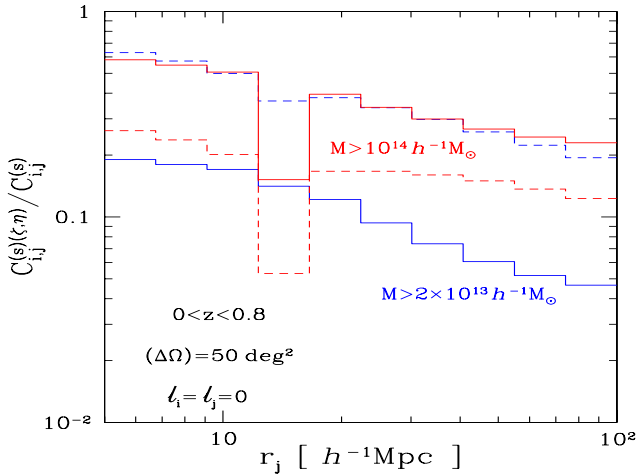}}
\end{center}
\caption{Relative importance of the non-Gaussian contributions
$C^{(s)(\zeta)}_{i,j}$ (solid line) and $C^{(s)(\eta)}_{i,j}$ (dashed line)
to the redshift-space covariance matrix $C^{(s)}_{i,j}$, along one row ($i=4$), 
associated with the distance bin $12.3<r<16.6 h^{-1}$Mpc.
We show the results obtained for halos in the redshift range
$0<z<0.8$ with an angular window of $50$ deg$^2$.}
\label{fig_Cij_NG}
\end{figure}

We show in Figs.~\ref{fig_Cii_NG} and \ref{fig_Cij_NG} the ratios
$C^{(s)(\zeta)}_{i,j}/C^{(s)}_{i,j}$ and $C^{(s)(\eta)}_{i,j}/C^{(s)}_{i,j}$.
They give the relative contribution to the covariance matrix of the terms associated
with the three-point correlation function, Eq.(\ref{Cij-zeta}), or with the four-point
correlation function, Eq.(\ref{Cij-eta}).
For low-mass halos the 4-pt contribution $C^{(s)(\eta)}_{i,j}$ is
larger than the 3-pt contribution $C^{(s)(\zeta)}_{i,j}$, whereas the 
ordering is reversed for high-mass halos. This agrees with the results obtained in
real-space in \citet{Valageas2011} (see their Figs.~15 and 16).
This behavior is due to the greater importance of shot-noise effects for rare
massive halos. This gives more weight to the 3-pt contribution, which is part of the
coupled shot-noise--sample-variance contribution $C^{(s)(3)}$ in Eq.(\ref{C3-def}), 
than to the 4-pt contribution, which is part of the pure sample-variance contribution 
$C^{(s)(4)}$ in Eq.(\ref{C4-def}).
For the same reason, these non-Gaussian contributions are relatively smaller
along the diagonal for rare halos, as seen on the fourth bin $j=i$ in 
Fig.~\ref{fig_Cij_NG}.

Figs.~\ref{fig_Cii_NG} and \ref{fig_Cij_NG} show that the non-Gaussian
contributions can make up to $20 - 60 \%$ of the full covariance matrix as soon
as one bin $i$ or $j$ corresponds to scales on the order of $10 h^{-1}$ Mpc.
Their relative contribution is also somewhat larger for off-diagonal terms.
Therefore, although these terms do not change the order of magnitude of the
covariance matrix, it is necessary to include them to obtain accurate estimates.

\subsubsection{Correlation matrices}
\label{correlations}

\begin{figure*}
\begin{center}
\epsfxsize=5.9 cm \epsfysize=5.9 cm {\epsfbox{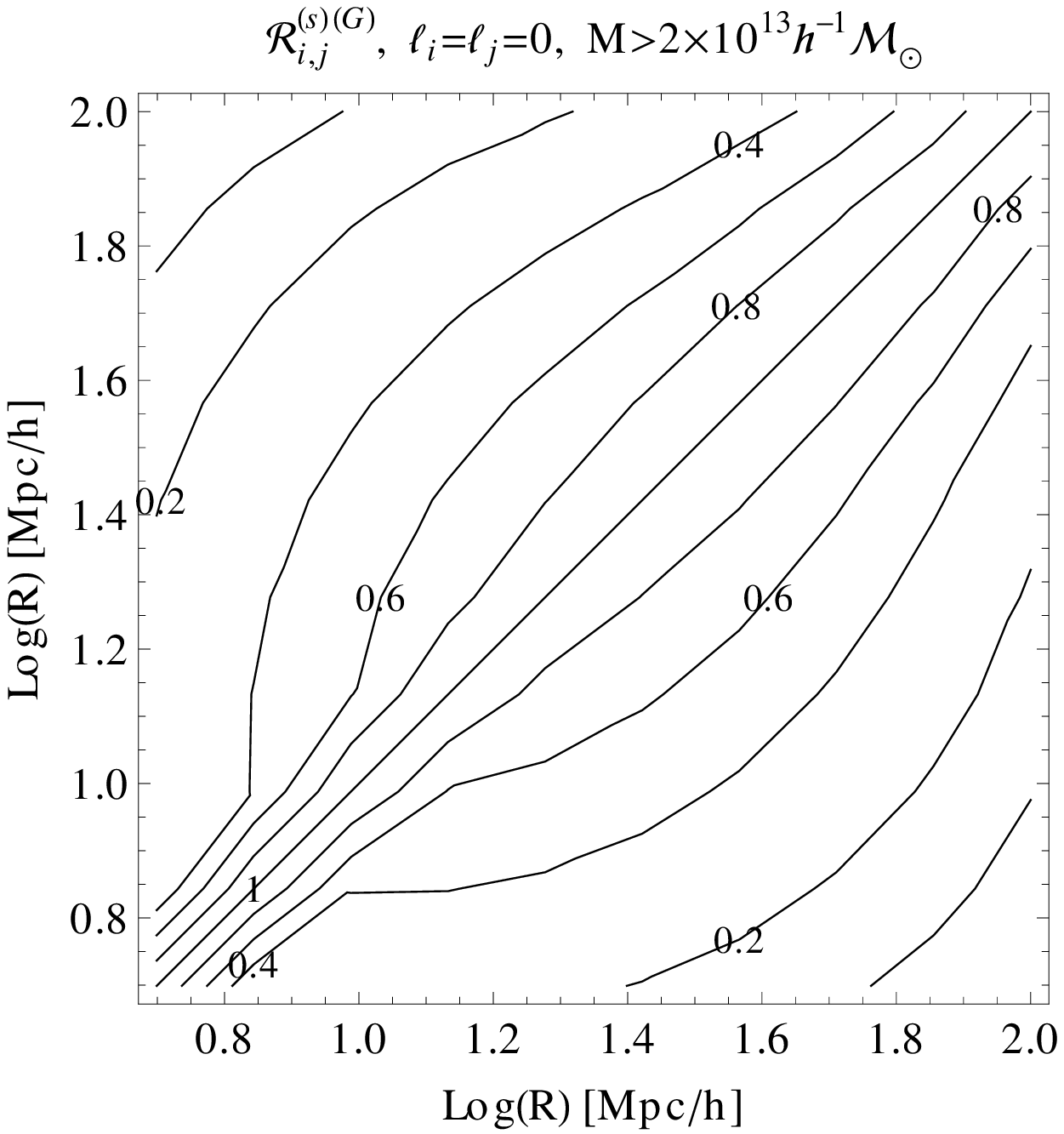}}
\epsfxsize=5.9 cm \epsfysize=5.9 cm {\epsfbox{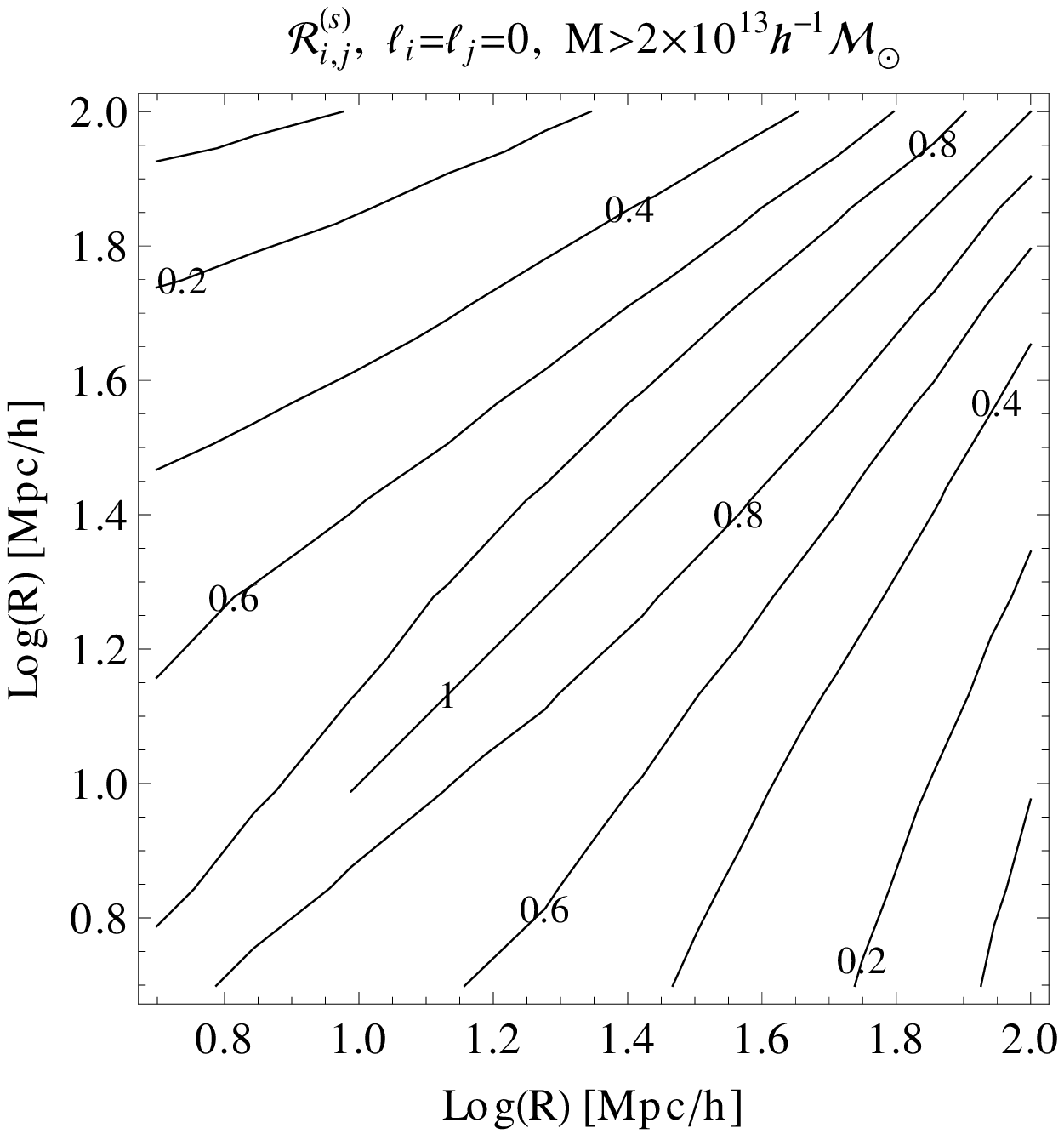}}\\
\epsfxsize=5.9 cm \epsfysize=5.9 cm {\epsfbox{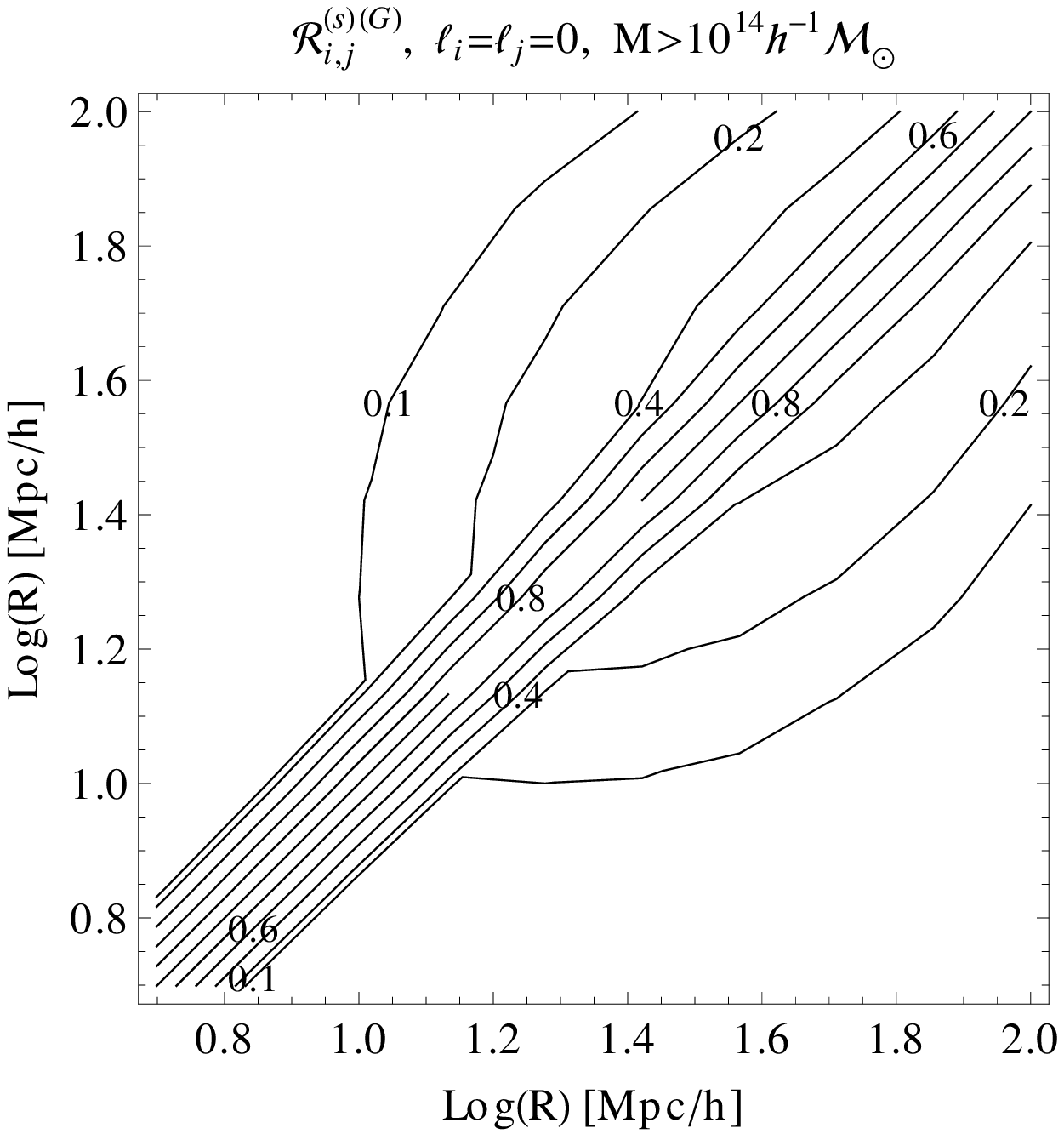}}
\epsfxsize=5.9 cm \epsfysize=5.9 cm {\epsfbox{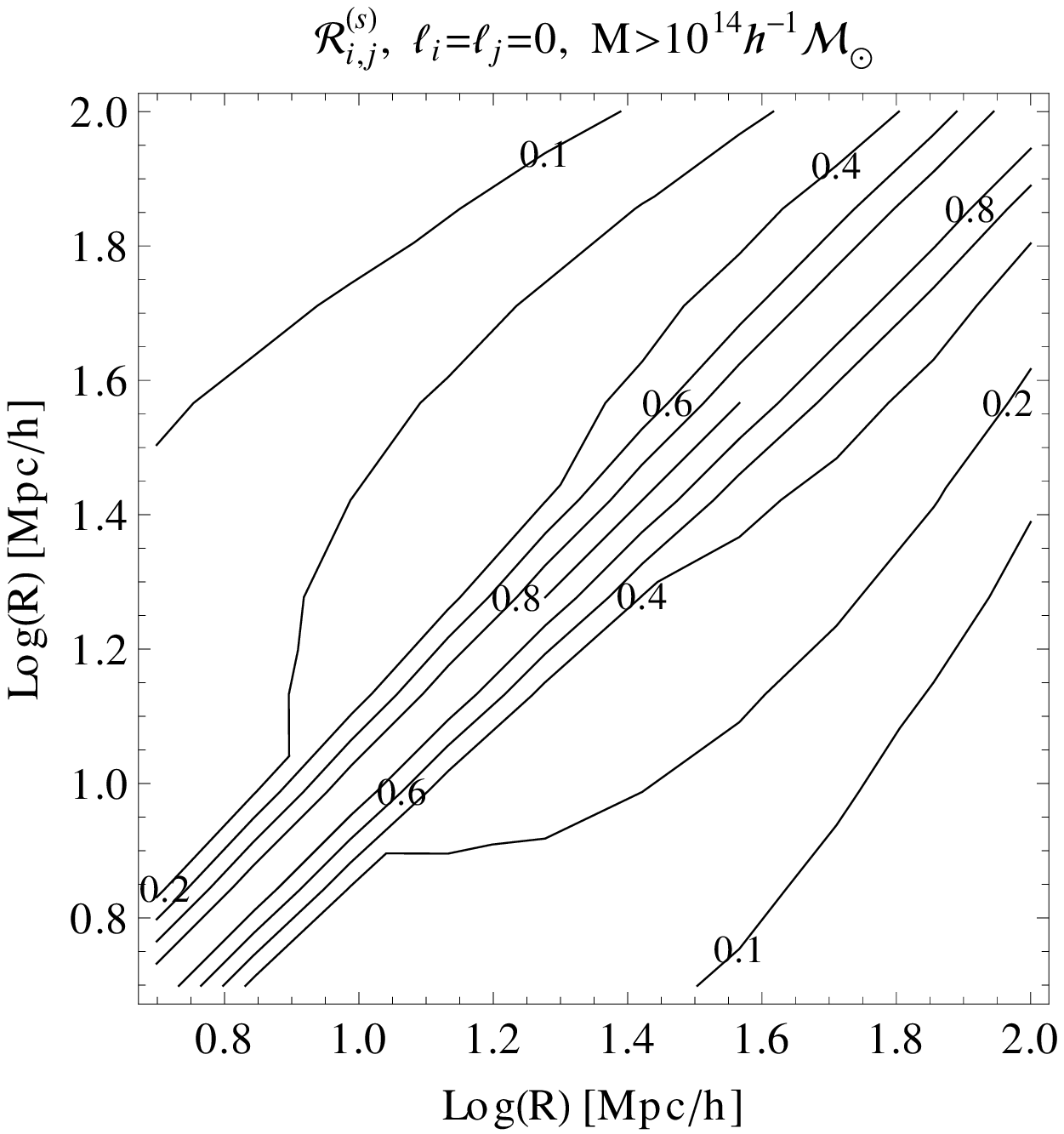}}
\end{center}
\caption{Contour plots for the redshift-space correlation matrices
$\cR^{(s)}_{i,j}$, defined as in Eq.(\ref{R-ij-def}). 
There are ten distance bins, over $5<r<100 h^{-1}$Mpc, equally spaced in
$\log(r)$, as in previous figures.
We considered halos in the redshift range $0<z<0.8$,
within an angular window of $50$ deg$^2$, above the mass thresholds
$M>2\times 10^{13}h^{-1} M_{\odot}$ in the {\it upper row}, and
$M>10^{14}h^{-1} M_{\odot}$ in the {\it lower row}.
{\it Left panels:} correlation matrix $\cR^{(s)(G)}_{i,j}$ associated with the
Gaussian part (\ref{Cij-G}) of the covariance matrix.
{\it Right panels:} full correlation matrix, associated with the full matrix
(\ref{Cij-G-NG}).}
\label{fig_Rij_CXis}
\end{figure*}

We show in Fig.~\ref{fig_Rij_CXis} the redshift-space correlation matrices defined
as
\beq
{\cal R}^{(s)}_{i,j} = \frac{C^{(s)}_{i,j}}{\sqrt{C^{(s)}_{i,i}C^{(s)}_{j,j}}} .
\label{R-ij-def}
\eeq
We plot the correlation matrices associated with the Gaussian part (\ref{Cij-G})
and with the full  matrix (\ref{Cij-G-NG}).
In agreement with Fig.~\ref{fig_Cij_NG}, we can see that the non-Gaussian
contributions, associated with the 3-pt and 4-pt correlations, make the
correlation matrix less diagonal, especially for low-mass halos.
As explained above, this is because rare massive halos show stronger pure
shot-noise effects that give an additional weight to the diagonal entries.
In any case, we can see that it is important to take into account non-Gaussian
contributions to obtain accurate estimates. Moreover, off-diagonal terms cannot
be neglected (especially for low-mass halos), hence it is necessary to use the
full covariance matrix $C^{(s)}_{i,j}$ to draw constraints on cosmology from
the analysis of X-ray cluster surveys.

\subsection{Covariance matrix of the ``$2\ell=2$'' redshift-space correlation}
\label{Covariance-2l=2}

We now study the covariance matrix $C^{(s)}_{i,j}$ of the ``$2\ell=2$''
redshift-space correlation, $\xi^{(s)}_{R_i,2}$, that is, $2\ell_i=2\ell_j=2$.

\begin{figure}
\begin{center}
\epsfxsize=8.5 cm \epsfysize=6.5 cm {\epsfbox{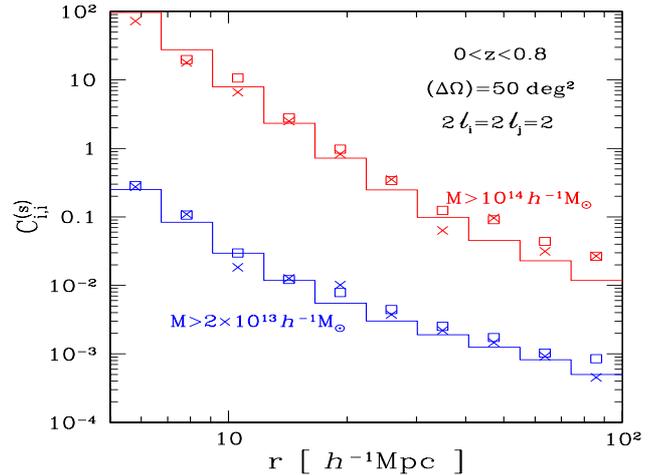}}
\end{center}
\caption{Covariance matrix $C^{(s)}_{i,j}$ along the 
diagonal, $i=j$, for the ``$2\ell=2$'' mutipole of the redshift-space correlation.
We show the results obtained for halos in the redshift range
$0<z<0.8$ with an angular window of $50$ deg$^2$.
Points from numerical simulations are obtained from 138 fields 
separated by a 10 degree gap (squares) or from 34 fields separated
by 20 degree gap (crosses).}
\label{fig_Cii_l1l1}
\end{figure}

\begin{figure}
\begin{center}
\epsfxsize=8.5 cm \epsfysize=6.5 cm {\epsfbox{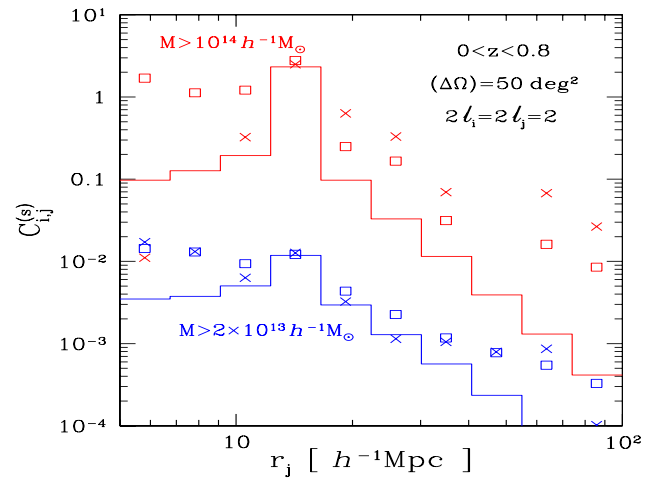}}
\end{center}
\caption{Covariance matrix $C^{(s)}_{i,j}$ as a
function of $j$, for $i=4$ associated with the distance bin
$12.3<r<16.6h^{-1}$Mpc, for the ``$2\ell=2$'' mutipole of the redshift-space 
correlation.
We show the results obtained for halos in the redshift range
$0<z<0.8$ with an angular window of $50$ deg$^2$.}
\label{fig_Cij_l1l1}
\end{figure}

\begin{figure}
\begin{center}
\epsfxsize=8.5 cm \epsfysize=6.5 cm {\epsfbox{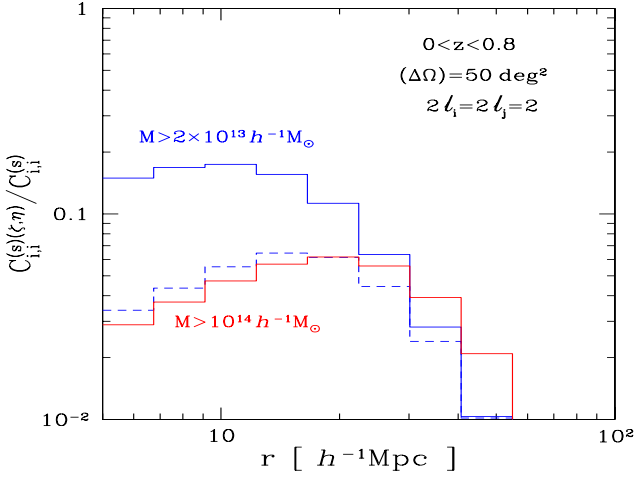}}
\end{center}
\caption{Relative importance of the non-Gaussian contributions
$C^{(s)(\zeta)}_{i,j}$ (solid line) and $C^{(s)(\eta)}_{i,j}$ (dashed line)
to the redshift-space covariance matrix $C^{(s)}_{i,j}$, along the diagonal, $i=j$.
We show the results obtained for halos in the redshift range
$0<z<0.8$ with an angular window of $50$ deg$^2$.
(The contribution $C^{(s)(\eta)}_{i,j}$ for $M>10^{14}h^{-1} M_{\odot}$ does not appear
because it is less than $10^{-2}$.)}
\label{fig_Cii_NG_l1l1}
\end{figure}

\begin{figure}
\begin{center}
\epsfxsize=8.5 cm \epsfysize=6.5 cm {\epsfbox{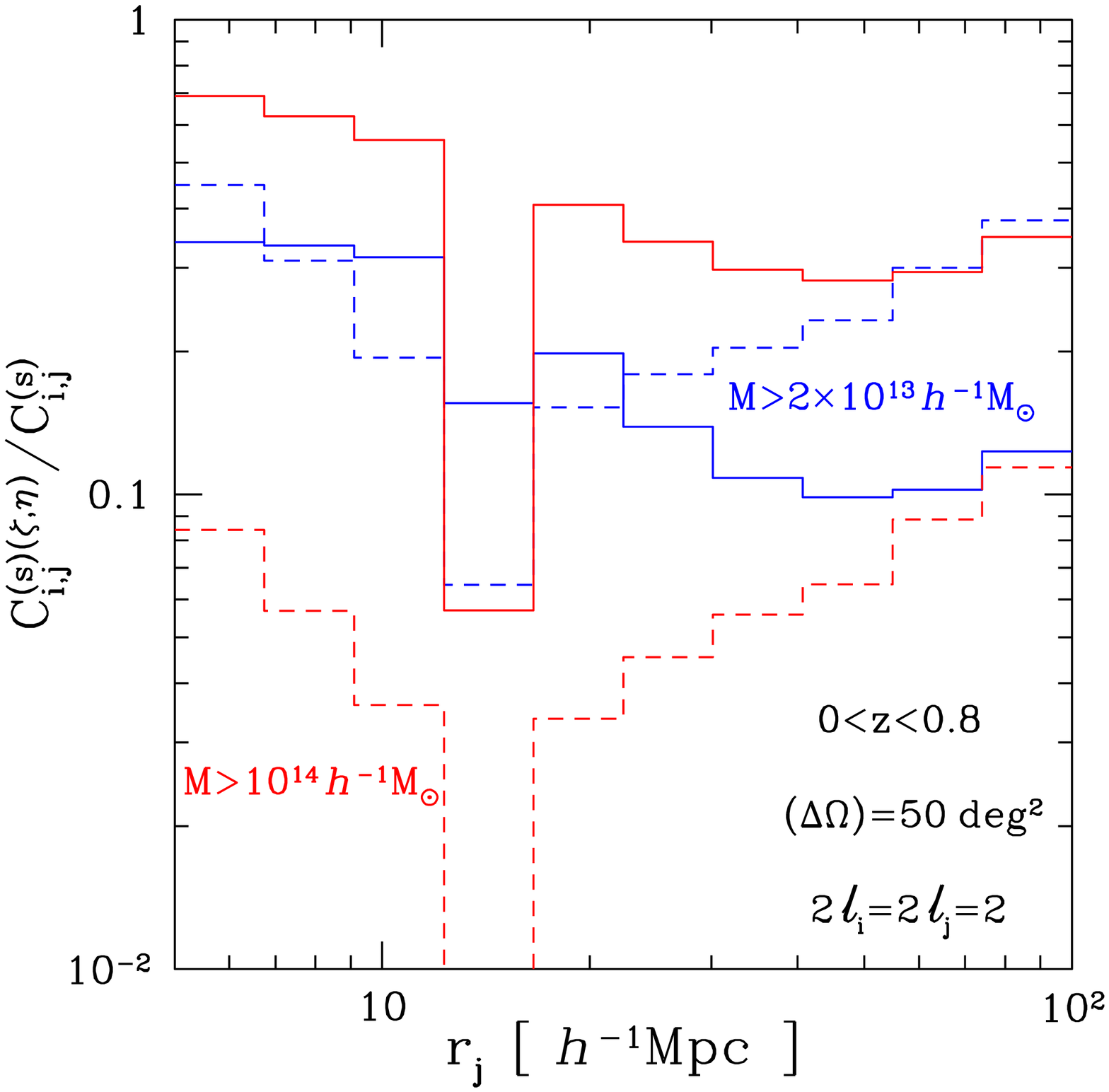}}
\end{center}
\caption{Relative importance of the non-Gaussian contributions
$C^{(s)(\zeta)}_{i,j}$ (solid line) and $C^{(s)(\eta)}_{i,j}$ (dashed line)
to the redshift-space covariance matrix $C^{(s)}_{i,j}$, along one row ($i=4$), 
associated with the distance bin $12.3<r<16.6 h^{-1}$Mpc.
We show the results obtained for halos in the redshift range
$0<z<0.8$ with an angular window of $50$ deg$^2$.}
\label{fig_Cij_NG_l1l1}
\end{figure}

We compare in Figs.~\ref{fig_Cii_l1l1} and \ref{fig_Cij_l1l1} our analytical
model with numerical simulations for the covariance matrix along the diagonal
and along one row. We considered two analyses of the numerical
simulations, where we used either 138 nonoverlapping fields separated by a
10 degree gap (squares) or 34 fields separated by a 20 degree gap (crosses).
This gives a measure of the statistical error of the simulation data, apart from
the systematic finite-resolution effects.
Along the diagonal (Fig.~\ref{fig_Cii_l1l1}) our predictions show reasonable 
agreement with the simulations, which appear reliable with low noise.
Along one row (Fig.~\ref{fig_Cij_l1l1}) the simulations give higher
off-diagonal entries than our model, but they also show a significant
statistical noise. In view of these uncertainties we obtain a reasonable
agreement on scales above $10 h^{-1}$Mpc. On smaller scales, our
model may underestimate the off-diagonal covariance because of nonlinear
effects, as suggested by the comparison for the mean correlation itself
shown in Fig.~\ref{fig_XisR_l1l1}.
Therefore, on scales above $10 h^{-1}$Mpc analytical models such as the
one presented in this paper can provide a competitive alternative to
numerical simulations to estimate quantities such as covariance matrices
that are difficult to obtain from numerical simulations (as they require
large box sizes with a good resolution and time-consuming data analysis
procedures).

We show in Figs.~\ref{fig_Cii_NG_l1l1} and \ref{fig_Cij_NG_l1l1} the
relative contribution to the covariance matrix of the terms associated with
the 3-pt and 4-pt correlation functions.
The comparison with Figs.~\ref{fig_Cii_NG} and \ref{fig_Cij_NG} 
shows that these non-Gaussian contributions are relatively less important
along the diagonal but more important on off-diagonal entries than for
the monopole ``$2\ell=0$'' case.
Thus, they make up to $40-60\%$ of the full covariance matrix over a large
range of scales.

For completeness we show in Appendix~\ref{Correlation-matrix-l=2}
the correlation matrices of this ``$2\ell=2$'' redshift-space correlation,
as well as the cross-correlation between the ``$2\ell=0$'' and ``$2\ell=2$''
components.
We find that the ``$2\ell=2$'' covariance matrix is significantly more
diagonal than the ``$2\ell=0$'' covariance matrix and that the cross-correlation
matrix between both components is quite small ($\sim 0.05$).
This means that the full covariance matrix in the $\{R_i,\ell_i\}$ space is
approximately block-diagonal (we may neglect entries with $\ell_i\neq\ell_j$).
Then, we can decouple the analysis of $\xi^{(s)}_{R_i,0}$ and 
$\xi^{(s)}_{R_i,2}$.

\section{Applications to real survey cases}
\label{Applications}

\begin{figure}
\begin{center}
\epsfxsize=8.5 cm \epsfysize=6.5 cm {\epsfbox{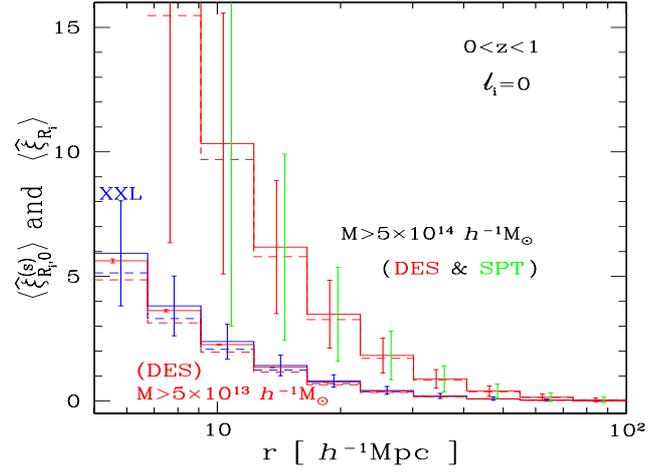}}
\end{center}
\caption{Mean monopole redshift-space (solid lines) and real-space (dashed lines) correlations,
$\lag\hxi^{(s)}_{R_i,0}\rag$ and $\lag\hxi_{R_i}\rag$, over ten comoving 
distance bins within $5<r<100 h^{-1}$Mpc, equally spaced in $\log(r)$.
We integrated over halos within the redshift
interval $0<z<1$, for the XXL, DES, and SPT surveys.
For DES we considered the mass thresholds $M>5\times 10^{13}h^{-1} M_{\odot}$
and $M>5\times 10^{14}h^{-1} M_{\odot}$ (smaller error bars), and for SPT
the mass threshold $M>5\times 10^{14}h^{-1} M_{\odot}$ (larger error bars
shifted to the right).
The error bars show the diagonal part of the redshift-space covariance
$\sigma_{\xis_i}=\sqrt{C_{i,i}^{(s)}}$.}
\label{fig_Xis_Xi_z0to1_XXLall}
\end{figure}

\begin{figure}
\begin{center}
\epsfxsize=8.5 cm \epsfysize=6.5 cm {\epsfbox{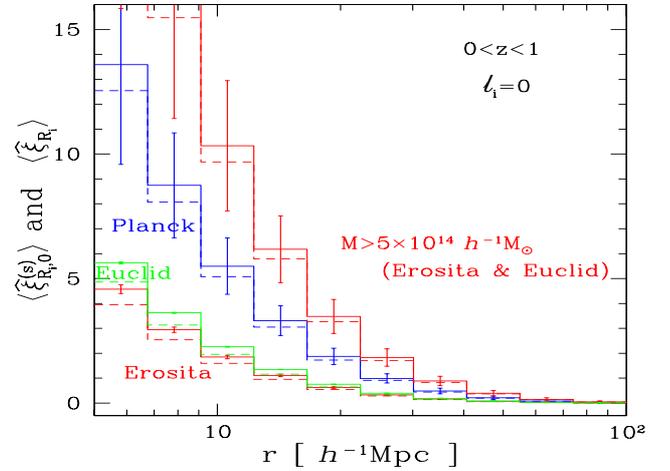}}
\end{center}
\caption{Mean monopole correlations, $\lag\hxi^{(s)}_{R_i,0}\rag$ (solid lines)
and $\lag\hxi_{R_i}\rag$ (dashed lines), 
as in Fig.~\ref{fig_Xis_Xi_z0to1_XXLall} but for all-sky surveys.
From top to bottom, we show a) halos above
$5\times 10^{14}h^{-1} M_{\odot}$ in either Erosita or Euclid, b) halos detected 
by Planck, c) halos above $5\times 10^{13}h^{-1} M_{\odot}$ in Euclid, and
d) halos detected by Erosita.}
\label{fig_Xis_Xi_z0to1_Planckall}
\end{figure}

\begin{figure}
\begin{center}
\epsfxsize=8.5 cm \epsfysize=6.5 cm {\epsfbox{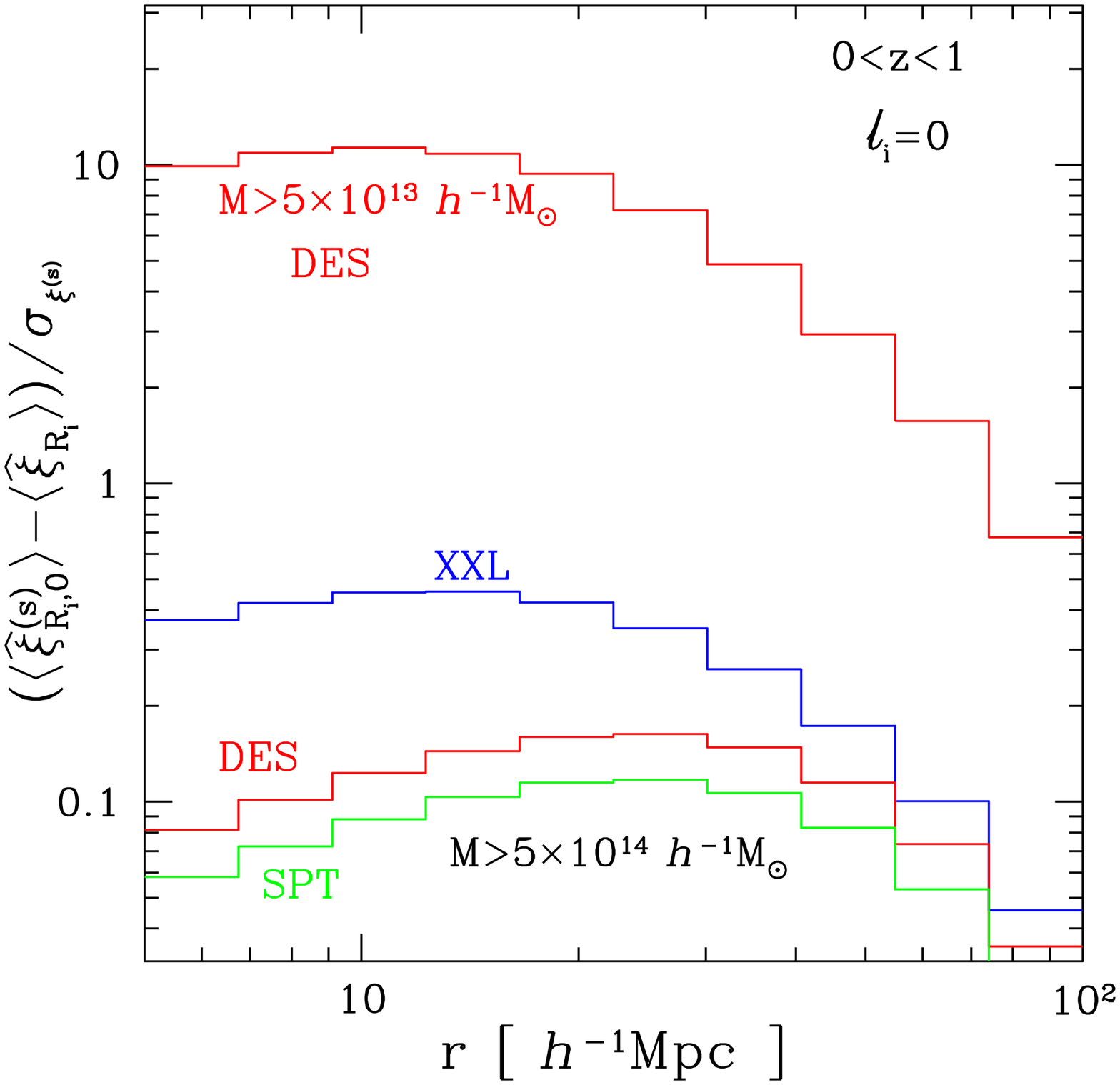}}
\end{center}
\caption{Ratio 
$(\lag\hxi^{(s)}_{R_i,0}\rag-\lag\hxi_{R_i}\rag)/\sigma_{\xis_i}$,
for the surveys shown in Fig.~\ref{fig_Xis_Xi_z0to1_XXLall}.}
\label{fig_Xis_Xi_sig_z0to1_XXLall}
\end{figure}

\begin{figure}
\begin{center}
\epsfxsize=8.5 cm \epsfysize=6.5 cm {\epsfbox{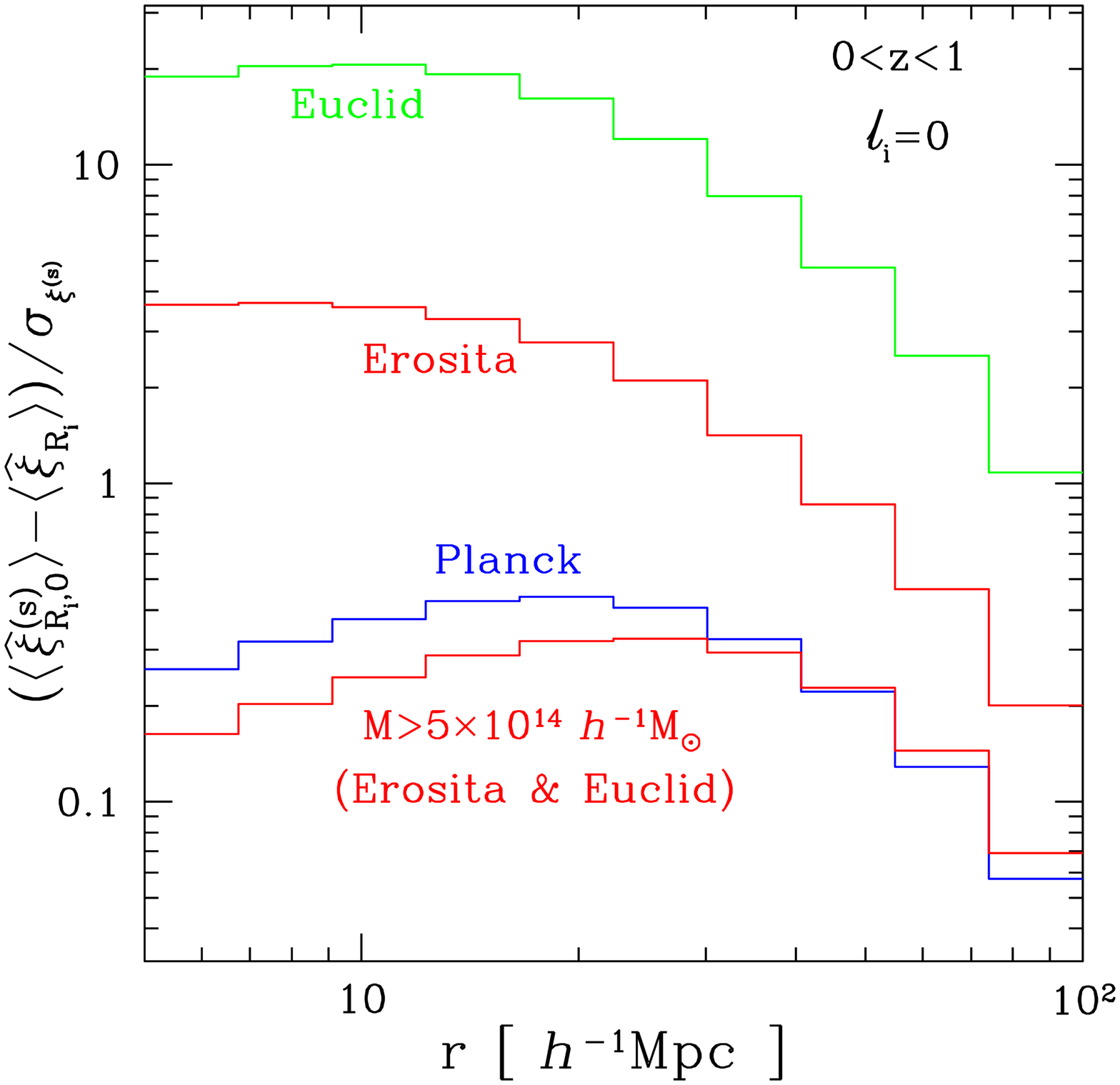}}
\end{center}
\caption{Ratio 
$(\lag\hxi^{(s)}_{R_i,0}\rag-\lag\hxi_{R_i}\rag)/\sigma_{\xis_i}$,
for the surveys shown in Fig.~\ref{fig_Xis_Xi_z0to1_Planckall}.}
\label{fig_Xis_Xi_sig_z0to1_Planckall}
\end{figure}

As in \citet{Valageas2011}, we now study the correlation functions that can be
measured in several cluster surveys.
We considered three surveys on limited angular windows:

- The XXL survey \citep{Pierre2011} is an XMM Very Large Programme 
specifically designed to constrain the equation of state of the dark energy 
by using clusters of galaxies. It consists of two $25$ \degs\ areas and 
probes massive clusters out to a redshift of $\sim 2$. We considered the ``C1
selection function'' given in Fig.J.1 of  \citet{Valageas2011}
(see also \citet{Pacaud2006,Pacaud2007}).

- The Dark Energy Survey (DES) is an optical imaging survey to cover $5,000$
\degs\ with the Blanco four-meter telescope at the Cerro Tololo Inter-American
Observatory\footnote{https://www.darkenergysurvey.org/index.shtml}.
We considered the expected mass threshold
$M>5\times 10^{13}h^{-1} M_{\odot}$, as well as the subset of massive 
clusters $M>5\times 10^{14}h^{-1} M_{\odot}$.

- The South Pole Telescope (SPT) operates at millimeter
wavelengths\footnote{http://pole.uchicago.edu/}. It  will cover some
$2,500$ \degs\ at three frequencies, aiming at detecting clusters of galaxies 
from the Sunyaev-Zel'dovich (S-Z) effect. We considered a mass threshold of
$5\times 10^{14}h^{-1} M_{\odot}$ \citep{Vanderlinde2010}.

We also considered three all-sky surveys.
In practice, the total angular area of these surveys is not really $4\pi$ sterad since
we must remove the galactic plane. In the following, for Planck we considered the
two-sided cone of angle $\theta_s=75$ deg (i.e., $|b|>15$ deg), which yields a total
area $\Delta\Omega\simeq 30576$ deg$^2$.
For Erosita and Euclid we took $\theta_s=59$ deg (i.e., $|b|>31$ deg), which
corresponds to a total area that is about one-half of the full sky,
$\Delta\Omega\simeq 20000$ deg$^2$.

- Planck operates at nine frequencies, enabling an efficient detection of the cluster 
S-Z signature but has a rather large PSF (5'-10'). Some 1625 massive clusters
out to $z=1$ are expected over the whole sky. We assumed the selection
function by \citet{Melin2006}, shown in Fig.J.1 of  \citet{Valageas2011}.

- For Erosita, a simple flux limit is currently  assumed as an average over the
whole sky: $4~10^{-14}$ \flux\ in the $[0.5-2]$ keV band \citep{Predehl2009}.
The associated selection function is shown in Fig.J.1 of  \citet{Valageas2011}.
This would yield $71,907$ clusters out to $z=1$. 

- For Euclid, we followed the prescription of the Euclid science book for the cluster
optical selection function and adopted a fixed mass threshold of 
$5\times 10^{13} h^{-1} M_{\odot}$ \citep{Refregier2010}.

In the following, the error bars shown in the figures are the statistical
error bars associated with the covariance matrices studied in
Sect.~\ref{Covariance-matrix} (i.e., we did not include other sources
of uncertainties such as observational noise). For the full-sky surveys
this ``sample-variance'' noise does not vanish because even full-sky surveys
up to a given redshift only cover a finite volume. In this limit the ``sample variance''
is usually called the ``cosmic variance''.

\subsection{Monopole redshift-space correlation}
\label{real-surveys-monopole}

We show the redshift-space and real-space correlations 
$\lag\hxi^{(s)}_{R_i,0}\rag$
and $\lag\hxi_i\rag$ that can be measured with these surveys in the redshift
range $0<z<1$ in Figs.~\ref{fig_Xis_Xi_z0to1_XXLall} and 
\ref{fig_Xis_Xi_z0to1_Planckall}. 
In each case, the redshift-space correlation is slightly higher than the
real-space correlation.
The error bars $\sigma_{\xis_i}=\sqrt{C_{i,i}^{(s)}}$ correspond to the
diagonal entries of the redshift-space covariance matrix.

Redshift-space distortions can be seen in these figures, but in several
cases they are smaller than the error bars.
To see the impact of these redshift-space distortions more clearly, we show
the ratio $(\lag\hxi^{(s)}_{R_i,0}\rag-\lag\hxi_{R_i}\rag)/\sigma_{\xis_i}$ in
Figs.~\ref{fig_Xis_Xi_sig_z0to1_XXLall} and \ref{fig_Xis_Xi_sig_z0to1_Planckall}.
This gives the difference between the monopole redshift-space and
real-space correlations in units of the standard deviation $\sigma_{\xis_i}$,
that is, of the error bar of the measure.

Among limited-area surveys, the amplification of the cluster correlation function
by the Kaiser effect will be clearly seen for halos above
$M>5\times 10^{13}h^{-1} M_{\odot}$ in DES. The effect is negligible for more
massive halos, $M>5\times 10^{14}h^{-1} M_{\odot}$, in DES and SPT, because
of their larger error bars (see Fig.~\ref{fig_Xis_Xi_z0to1_XXLall}).
For the ``C1 clusters'' measured in the XXL survey redshift distortions are marginally
relevant: they are on the order of $0.4  \sigma_{\xis_i}$ in distance bins
around $15 h^{-1}$Mpc, so that by collecting the signal from all distance bins
they would give a signal-to-noise ratio of about unity.

Among full-sky surveys, the redshift-space amplification will have a clear impact
for the full cluster populations measured in Erosita and Euclid, and a marginal
impact for Planck.
Again, subsamples of massive clusters, $M>5\times 10^{14}h^{-1} M_{\odot}$, 
lead to large error bars that hide this redshift-space effect.

In practice one does not measure both real-space and redshift-space
correlations from a cluster survey, because one does not have a map of the
cluster velocities. Therefore, the redshift-space amplification of the spherically
averaged correlation $\lag\hxi^{(s)}_{R_i,0}\rag$ is degenerate with the halo
bias $\bb$ (which is difficult to predict with a high accuracy).
To measure redshift-space distortions, one needs to measure the anisotropies
of the redshift-space correlation, that is, the higher-order multipoles
$\xi^{(s;2)}$ and $\xi^{(s;4)}$ of Eq.(\ref{xis}).
This provides a distinctive signature that breaks the degeneracy with the
halo bias and yields another constraint on cosmology through the factor $\beta$,
that is, the growth rate $f$ of the density fluctuations defined in Eq.(\ref{f-def}).
We consider these higher-order multipoles in the following sections.

\subsection{``$2\ell=2$'' redshift-space correlation}
\label{real-surveys-2l=2}

\begin{figure}
\begin{center}
\epsfxsize=8.5 cm \epsfysize=6.5 cm {\epsfbox{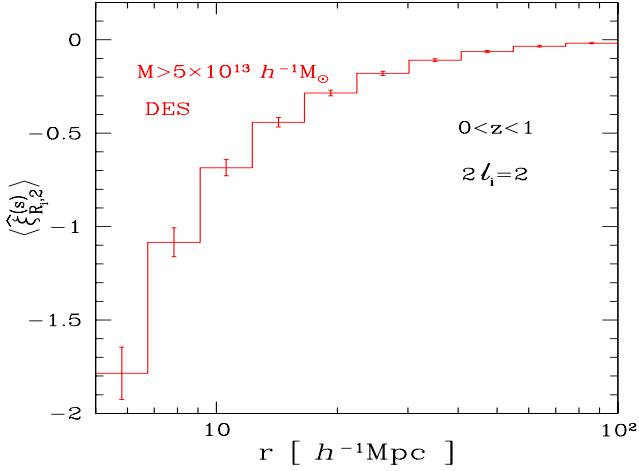}}
\end{center}
\caption{Mean ``$2\ell=2$'' redshift-space correlation 
$\lag\hxi^{(s)}_{R_i,2}\rag$. As in Fig.~\ref{fig_Xis_Xi_z0to1_XXLall}, we considered
ten comoving distance bins within $5<r<100 h^{-1}$Mpc, equally spaced in
$\log(r)$, and integrated over halos within the redshift interval $0<z<1$.
We show our results for halos above $M>5\times 10^{13}h^{-1} M_{\odot}$
for DES. 
The error bars show the diagonal part of the redshift-space covariance
$\sigma_{\xis_i}=\sqrt{C_{i,i}^{(s)}}$.}
\label{fig_Xis_l1l1_z0to1_XXLall}
\end{figure}

\begin{figure}
\begin{center}
\epsfxsize=8.5 cm \epsfysize=6.5 cm {\epsfbox{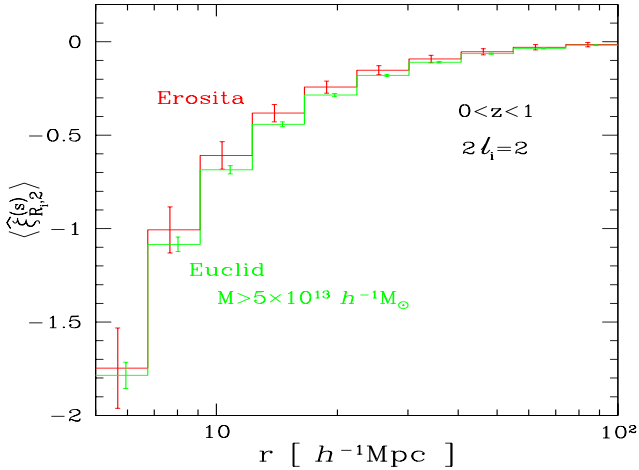}}
\end{center}
\caption{Mean ``$2\ell=2$'' redshift-space correlation 
$\lag\hxi^{(s)}_{R_i,2}\rag$, as in Fig.~\ref{fig_Xis_l1l1_z0to1_XXLall}
but for all-sky surveys.
We show our results for all halos detected by Erosita (upper curve) and for halos
above $5\times 10^{13}h^{-1} M_{\odot}$ in Euclid (lower curve).}
\label{fig_Xis_l1l1_z0to1_Planckall}
\end{figure}

We show in Figs.~\ref{fig_Xis_l1l1_z0to1_XXLall} and 
\ref{fig_Xis_l1l1_z0to1_Planckall} the mean ``$2\ell=2$'' redshift-space
correlation $\lag\hxi^{(s)}_{R_i,2}\rag$. We considered the same surveys
as in Sect.~\ref{real-surveys-monopole}, but we only plot our results for
the cases where the signal-to-noise ratio is higher than unity.
In agreement with the analysis of Sect.~\ref{Two-point-correlation},
this quadrupole term is more difficult to measure than the monopole
term studied in Sect.~\ref{real-surveys-monopole}, and only DES, Erosita,
and Euclid can obtain a clear detection. Even for these surveys, the correlation
can only be measured for the full halo population and error bars become too
large if one restricts to the subsample of rare massive halos above
$5\times 10^{14}h^{-1} M_{\odot}$.
Nevertheless, it will be useful to measure this ``$2\ell=2$'' redshift-space
correlation in these surveys because this should tighten the constraints on
cosmology. Indeed, a simultaneous analysis of $\lag\hxi^{(s)}_{R_i,0}\rag$
and $\lag\hxi^{(s)}_{R_i,2}\rag$ constrains the factor $\beta=f/\bb$, and in turn
the halo bias and the growth rate of density fluctuations.

\subsection{``$2\ell=4$'' redshift-space correlation}
\label{real-surveys-2l=4}

\begin{figure}
\begin{center}
\epsfxsize=8.5 cm \epsfysize=6.5 cm {\epsfbox{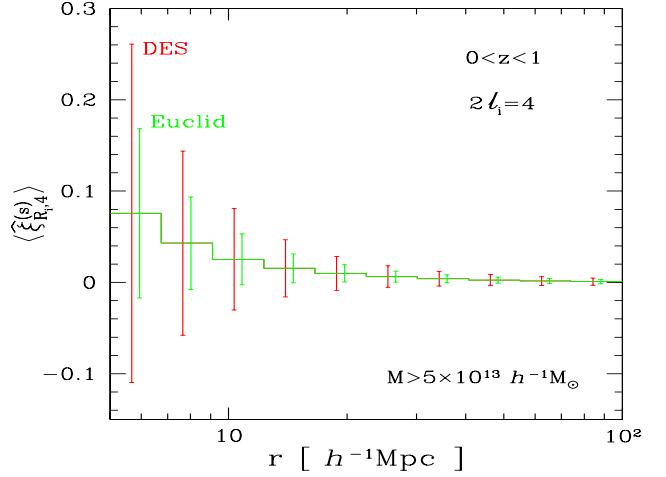}}
\end{center}
\caption{Mean ``$2\ell=4$''redshift-space correlation
$\lag\hxi^{(s)}_{R_i,4}\rag$. As in Fig.~\ref{fig_Xis_Xi_z0to1_XXLall}, we considered
ten comoving distance bins within $5<r<100 h^{-1}$Mpc, equally spaced in
$\log(r)$, and integrated over halos within the redshift interval $0<z<1$.
We show our results for halos above $M>5\times 10^{13}h^{-1} M_{\odot}$
for DES (left error bars) and Euclid (right error bars).}
\label{fig_Xis_l2l2_z0to1_all}
\end{figure}

We show in Fig.~\ref{fig_Xis_l2l2_z0to1_all} the mean ``$2\ell=4$'' redshift-space 
correlation $\lag\hxi^{(s)}_{R_i,4}\rag$.
The signal is even weaker than for the ``$2\ell=2$'' correlation and only
DES and Euclid may be able to obtain a detection.
DES is unlikely to provide an accurate measure but it should still give useful
upper bounds.
Euclid may obtain a significant measure of the overall amplitude
by gathering the information from all distance bins. This would provide
an additional constraint on $\beta$ to the one provided by the
mutipole ``$2\ell=2$'' of Fig.~\ref{fig_Xis_l1l1_z0to1_Planckall}.

\section{Conclusion}
\label{Conclusion}

We have generalized our previous study of the real-space
correlation functions of clusters of galaxies \citep{Valageas2011} to
include redshift-space distortions.
On large scales they lead to an anisotropic correlation function
because of the Kaiser effect, due to the correlation between density
and velocity fields. There are no ``fingers of God'' because clusters are the
largest bound objects, as opposed to galaxies that can have performed
several orbital revolutions within a larger halo and show a high small-scale
velocity dispersion.
Then, taking into account this Kaiser effect at leading order and using
a simple hierarchical model for the three-point and four-point halo correlations,
we developed an analytical formalism to obtain explicit expressions
for the mean redshift-space correlations and their covariance matrices.
We included shot-noise and sample-variance effects as well as Gaussian
and non-Gaussian contributions.
This is a direct extension of the formalism presented in \citet{Valageas2011}
for real-space correlations.

Expanding as usual the angular anisotropy of the redshift-space correlation
on Legendre polynomials, we considered the first three nonzero multipoles:
the monopole $2\ell=0$ (i.e. the spherically-averaged correlation), the
quadrupole, $2\ell=2$, and the next multipole, $2\ell=4$.
Higher-order multipoles should arise from nonlinear corrections to the Kaiser
effect. Although they are not studied here, which would require a more complex
nonlinear modeling of the Kaiser effect itself, their analysis could be performed 
using the formalism developped in this paper (one only needs to include
higher multipoles in the sums that appear in the expressions of the
covariance matrices).
However, for practical purposes they are probably irrelevant for clusters
of galaxies. Indeed, higher-order multipoles are increasingly difficult to measure
and we found that the third mutipole, $2\ell=4$, may only be detected
(at a $1-\sigma$ level) by Euclid or a similar survey.

We obtained a reasonable agreement with numerical simulations for the mean
correlations and covariance matrices on large scales ($r> 10 h^{-1}$Mpc).
Redshift-space distortions amplify the monopole correlation by about
$10-20\%$, depending on the halo mass. The covariance matrix is also
amplified by $10-40\%$ and the signal-to-noise ratio remains of the same order
as for the real-space correlation.
We found that non-Gaussian terms can contribute to the covariance matrices
up to $20-60\%$, especially for nondiagonal entries when one of the two
bins corresponds to small scales ($\sim 10 h^{-1}$Mpc).
They make the correlation matrices less diagonal.
We also found that the correlation matrices of the quadrupole ($2\ell=2$)
are significantly more diagonal than the correlation matrices of the monopole
($2\ell=0$).
On the other hand, the cross-correlations of the monopole and quadrupole
are quite small, on the order of $0.05$. This means that the full covariance
matrix is approximately block-diagonal in the space $\{R_i,2\ell_i\}$, that is,
we may decouple the analysis of the monopole and quadrupole.

Our predictions for ongoing and future surveys (only taking into account the
sample-variance and shot-noise contributions) show that the impact of
redshift distortions on the spherical-averaged correlation (i.e. the monopole)
will only be significant for DES, Erosita, and Euclid.
For the other surveys investigated here, XXL, SPT, and Planck, the difference
$\lag\hxi^{(s)}_{R_i,0}\rag - \lag\hxi_{R_i}\rag$, between the redshift-space
and real-space correlations, falls below ``$1-\sigma$'' of the statistical error
bar.
In practice, even for these surveys it would be better to compare the data
with theoretical predictions for the redshift-space correlations rather than the
real-space correlations, since simple models for the latter are not more difficult
to implement and one does not need a very accurate modeling of redshift 
distortions. For covariance matrices, it is sufficient to use the real-space
ones, which are easier to compute.
For DES, Erosita, and Euclid, one should take into account redshift-space
distortions for both the mean correlations and their covariance matrices.
However, because of their modest impact on the covariance matrices
(see Figs.~\ref{fig_Cii} and \ref{fig_Cij}) and the limited accuracy of such
computations, it is probably sufficient to use the real-space covariance matrices
in a first step.

The quadrupole, $2\ell=2$, is more difficult to measure from the distribution
of clusters of galaxies, and we found that only DES, Erosita, and Euclid, among
the surveys considered here, will likely provide a useful measure.
In this respect, clusters are less efficient than galaxies in measuring
redshift-space distortions. Even though clusters have a higher bias and
stronger correlations, they are much less numerous than galaxies. This leads 
to a lower signal-to-noise ratio for the quadrupole.
Nevertheless, such a measure would provide a nice supplement and would
help tighten the constraints on cosmology.
An extension of this paper would be to study on a quantitative level the
expected improvement brought by this measure for these surveys.
This requires a statistical analysis, using Fisher matrices or likelihood functions,
which we leave for future works.

The last multipole generated at linear order by the Kaiser effect,
$2\ell=4$, is very difficult to measure from cluster surveys and we found
that only Euclid will likely be able to provide a meaningful measure.
Nonlinear corrections will generate higher-order multipoles but they are
likely to be even more difficult to measure. Therefore, for practical purposes
redshift-space distortions should only impact cluster correlations at the level of
the monopole and quadrupole components.

Our results deliberately disregard observational constraints related to cluster redshifts estimation. In practice, measuring cluster redshifts is a costly task that consists in obtaining spectra for a few ($> 10$) galaxies pertaining to each cluster, leading to errors on the order of
$\sigma(z) < 0.01 (1+z)$. This precision depends on the number of measured spectra, which in turn depends on cluster richness (and consequently, on its mass), cluster apparent size, morphological complexity, dynamical state, and distance to the observer, as well as on fixed instrumental constraints.
In some cases, only the brightest central galaxy (BCG) can have its redshift measured. This may result in a biased cluster redshift estimate, for the BCG may have a small proper motion within the cluster potential well \citep{Adami2000, Oegerle2001, Coziol2009}.
In many other cases, only photometric redshifts will be obtained, resulting in larger redshift uncertainties -- on the order of $\sigma(z) \sim 0.05 (1+z)$ \citep[e.g.][]{Pillepich2012}. However, recent studies have underlined the potential of photometric surveys for studying redshift space distortions at the galaxy scale, using the information in the (2D) angular correlation function \citep[e.g.][]{Ross2011}. We leave the application of this method to clusters to future work, and note that a quantitative analysis (e.g.~a Fisher analysis) would largely benefit from a proper modeling of follow-up incompleteness, uncertainties, and biases on cluster redshifts given the halo masses and positions.

\acknowledgements{We thank M.~Pierre and C.~Adami for useful discussions. We thank R.~Teyssier for providing us with the halo catalog from the ``horizon'' simulation. This work was performed using HPC resources from GENCI-CCRT (Grant 2012-t2012046803).}

\appendix

\section{Computation of geometrical averages}
\label{geometrical-averages}

We give in this appendix the explicit expressions of the various geometrical
averages of two-point correlations that are needed to compute the covariance
matrix $C^{(s)}_{i,j}$, within our framework. We followed the method and the
notations of \citet{Valageas2011} while we extend their results to redshift-space
(especially their Appendix F).

The second term in
Eq.(\ref{Cij-G}) involves the average (\ref{I3-ij-xi-def}) of the correlation between
two concentric spherical shells.
Defining the 3D Fourier transform of the $2\ell$-multipole top-hat as
\beq
\tW_3^{(2\ell)}(\vk R) = \int_0^R \!\! \frac{\dd \vr}{4\pi R^3/3} \, (4\ell \!+\! 1)
P_{2\ell}(\mu_{\vr}) \, e^{\ii\vk\cdot\vr} ,
\label{W3-l-def}
\eeq
where again $\mu_{\vr}=(\ve_z\cdot\vr)/r$,
the 3D Fourier-space window of the $i$-shell reads as
\beqa
\tW^{(2\ell)}_i(\vk) & = & \int_{\cV_i}\frac{\dd\vr}{\cV_i} \, (4\ell \!+\! 1)
P_{2\ell}(\mu_{\vr}) \, e^{\ii\vk\cdot\vr}
\nonumber \\
& = & \frac{\Rip^3 \tW_3^{(2\ell)}(\vk\Rip) - \Rim^3 \tW_3^{(2\ell)}(\vk\Rim)}
{\Rip^3- \Rim^3} .
\label{W3-D-def}
\eeqa
In Eq.(\ref{W3-l-def}) the subscript $3$ recalls that we consider a 3D
window. To simplify notations we did not recall this fact in the window
$\tW^{(2\ell)}_i$ associated with each 3D radial bin.
In studies of the real-space correlation \citep{Valageas2011},
only the usual top-hat window $\tW_3^{(0)}(k R)$ appears, which does not
depend on the angle of $\vk$. Here, because we study the anisotropic
redshift-space correlation and its various multipoles (\ref{xi-3}) higher-order
multipoles $\tW_3^{(2\ell)}(\vk R)$ appear, which depend on the direction
of $\vk$.
Eq.(\ref{W3-l-def}) also reads as
\beq
\tW_3^{(2\ell)}(\vk R) = \hW_3^{(2\ell)}(k R) \; P_{2\ell}(\mu_{\vk}) ,
\eeq
where the angular dependence has been factored out, with
\beq
\hW_3^{(2\ell)}(k R) = (-1)^{\ell} \, (4\ell \!+\! 1) \int_0^R \frac{\dd r \, 3r^2}{R^3} \;
j_{2\ell}(k r) .
\label{Whi-def}
\eeq
This yields for the window of the $i$-shell,
\beq
\tW_i^{(2\ell)}(\vk) = \hW_i^{(2\ell)}(k) \; P_{2\ell}(\mu_k) ,
\eeq
with
\beqa
\hW^{(2\ell)}_i(k) = 
\frac{\Rip^3 \hW_3^{(2\ell)}(k\Rip) - \Rim^3 \hW_3^{(2\ell)}(k\Rim)}
{\Rip^3- \Rim^3} .
\label{hW3-D-def}
\eeqa
In practice, the redshift-space correlation of clusters of galaxies should be
dominated by the Kaiser effect and the lower multipoles $2\ell=0, 2$, and $4$
as in Eqs.(\ref{xis})-(\ref{xis-4}). Therefore, we only give explicit expressions
for low-order multipoles, which read as
\beq
\hW_3^{(0)}(k R) = 3 \, \frac{\sin(kR)-kR\cos(kR)}{(kR)^3} ,
\label{W3-0}
\eeq
\beqa
\hW_3^{(2)}(k R) & = & \frac{-15}{(kR)^3}  \left[ kR\cos(kR) - 4 \sin(kR)
+ 3 \, {\rm Si}(kR) \right] , \nonumber \\
&&
\label{W3-2}
\eeqa
\beqa
\hW_3^{(4)}(k R) & = & \frac{27}{2(kR)^5}  \left[ (105 \!-\! 2(kR)^2) kR\cos(kR)
\right. \nonumber \\
&& \hspace{-1cm} \left. - (105 \!-\! 22(kR)^2) \sin(kR) + 15 (kR)^2 \, 
{\rm Si}(kR) \right] ,
\label{W3-4}
\eeqa
where ${\rm Si}(z)= \int_0^z \dd t \, \sin(t)/t$ is the sine integral.
Then, writing the two-point correlation function in Eq.(\ref{I3-ij-xi-def})
in terms of the power spectrum (\ref{Ps}), the angular integration gives
\beqa
\overline{\xis_{i',j'}} & = & \int_0^{\infty} \frac{\dd k}{k} \, 
\Delta^2(k,z) \, \hW_i^{(2\ell_i)}(k) \, \hW_j^{(2\ell_j)}(k)  \nonumber \\ 
&& \hspace{-0.4cm} \times \left\{  \left(1\!+\!\frac{2\beta}{3}\!+\!\frac{\beta^2}{5}\right)
\left( \bea{ccc} 0 & 2\ell_i & 2\ell_j  \\ 0 & 0 & 0 \ea \right)^{\!\!2} 
+ \left(\frac{4\beta}{3}\!+\!\frac{4\beta^2}{7}\right) \right. \nonumber \\
&& \hspace{-0.4cm} \left. \times 
\left( \bea{ccc} 2 & 2\ell_i & 2\ell_j  \\ 0 & 0 & 0 \ea \right)^{\!\!2}
+ \frac{8\beta^2}{35}
\left( \bea{ccc} 4 & 2\ell_i & 2\ell_j  \\ 0 & 0 & 0 \ea \right)^{\!\!2} \right\} .
\label{I3-ij-def}
\eeqa

Next, the last term in Eq.(\ref{Cij-G}) involves the non-connected four-point average
\beqa
\overline{\xis_{i;j'} \xis_{i';j}} & = & \int \frac{\dd\chi_j}{\cD_i} 
\int\frac{\dd\vOm_i\dd\vOm_j}{(\Delta\Omega)^2} 
\int_i\frac{\dd\vs_{i'}}{\cV_i} \int_j\frac{\dd\vs_{j'}}{\cV_j} (4\ell_i \!+\! 1)
\nonumber \\
&& \times (4\ell_j \!+\! 1) P_{2\ell_i}(\mu_{i'}) P_{2\ell_j}(\mu_{j'}) \; \xis_{i;j'} \xis_{i';j} .
\label{Kij-def}
\eeqa
As in \citet{Valageas2011}, expressing the two-point correlation functions in terms
of the power spectrum, using the flat-sky (small angle) approximation, as well
as Limber's approximation \citep{Limber1953,Kaiser1992,Munshi2008},
we obtain
\beqa
\overline{\xis_{i;j'} \xis_{i';j}}  & = & \frac{2\pi}{\cD} \int \dd\vk_1\dd\vk_2 \,
P(k_1) (1\!+\!\beta\mu_1^2)^2 \; P(k_2) (1\!+\!\beta\mu_2^2)^2 \nonumber \\ 
&& \times \, \delta_D(k_{1\parallel}+k_{2\parallel}) \tW_i^{(2\ell_i)}(\vk_1) \, 
\tW_j^{(2\ell_j)}(\vk_2) \nonumber \\ 
&& \times \, \tW_2(|\vk_{1\perp}+\vk_{2\perp}|\cD\theta_s)^2 ,
\label{xi-ij'-i'j-1}
\eeqa
where $\mu_i=(\ve_z\cdot\vk_i)/k_i$, $\theta_s$ is the angular
radius of the survey window (in the flat-sky approximation),
\beq
(\Delta\Omega) = \pi \theta_s^2 ,
\label{thetas-def}
\eeq 
and $\tW_2$ is the 2D Fourier-space circular window,
\beq
\tW_2(k_{\perp}\cD\theta_s) = \int \frac{\dd\vtheta}{\pi\theta_s^2} \,
e^{\ii\vk_{\perp}\cdot\cD\vtheta} =
\frac{2J_1(k_{\perp}\cD\theta_s)}{k_{\perp}\cD\theta_s} .
\label{tW2}
\eeq
Introducing an auxiliary wavenumber $\vk_{\perp}$ with a Dirac factor
$\int\dd\vk_{\perp}\delta_D(\vk_{\perp}-\vk_{1\perp}-\vk_{2\perp})$
and using the exponential representation of Dirac functions, we can partially
factorize the integrals and perform a few angular integrations.
This yields
\beq
\overline{\xis_{i;j'} \xis_{i';j}} = \int \frac{\dd \vr}{2\pi r \cD^2\theta_s}
\; \cI_{i,2\ell_i}^{(s)}(\vr) \, \cI_{j,2\ell_j}^{(s)}(\vr) \, 
A^{(s)}\!\left(\!\frac{\vr}{\cD\theta_s}\!\right) ,
\label{Kij-3}
\eeq
where we introduced
\beq
A^{(s)}\!\left(\!\frac{\vr}{\cD\theta_s}\!\right) = \frac{r\cD\theta_s}{2\pi} 
\int \dd \vk_{\perp} \; e^{\ii\vr_{\perp}\cdot\vk_{\perp}} \, 
\tW_2(k_{\perp}\cD\theta_s)^2 
\label{A3s_r-def}
\eeq
and
\beq
\cI_{i.2\ell_i}^{(s)}(\vr) = \int \dd\vk \, e^{\ii\vk\cdot\vr} \, P(k) 
(1+\beta\mu_{\vk}^2)^2 \, \tW_i^{(2\ell_i)}(\vk) .
\label{cIs-def}
\eeq
Using the multipole decomposition (\ref{xis}) and the expansion of plane
waves on spherical harmonics, we can write $\cI_{i,2\ell_i}^{(s)}(\vr)$
as
\beq
\cI_{i,2\ell_i}^{(s)}(\vr) = \sum_{\ell=0}^4 \cI_{i,2\ell_i}^{(s;2\ell)}(r) 
\, P_{2\ell}(\mu_{\vr})  ,
\label{cIs-n}
\eeq
where the sum over $\ell$ runs up to $4$ because the expansion (\ref{xis})
only includes the terms $0\leq\ell\leq 2$ and we only study estimators
$\hxi^{(s)}_{R_i,2\ell_i}$ with $0\leq\ell_i\leq 2$.
If we consider a model where the redshift-space correlation includes multipoles
up to $2\ell_{\rm max}$ and we study estimators up to order $2\hat{\ell}_{\rm max}$
the sum (\ref{cIs-n}) runs up to $\ell\leq \ell_{\rm max}+\hat{\ell}_{\rm max}$.
Here we introduced
\beqa
\cI_{i,2\ell_i}^{(s;2\ell)}(r) & = & (4\ell \!+\! 1) \sum_{\ell'=0}^2 \int_0^{\infty} 
\frac{\dd x \; x \, \xi^{(s;2\ell')}(x)} {r(\Rip^3-\Rim^3)} \nonumber \\
&& \times \left[ \Rip^2 \cW_{\ell_i}^{(\ell,\ell')}\!
\left(\frac{r}{\Rip},\frac{x}{\Rip}\right) \right. \nonumber \\
&& \left.  - \Rim^2 \cW_{\ell_i}^{(\ell,\ell')} \!
\left(\frac{r}{\Rim},\frac{x}{\Rim}\right) \right] ,
\label{cI-3-W3}
\eeqa
and
\beqa
\cW_{\ell_i}^{(\ell,\ell')}(a,b) & = & (-1)^{\ell+\ell'} 
\left( \bea{ccc} 2\ell & 2\ell' & 2\ell_i  \\ 0 & 0 & 0 \ea \right)^{\!\!2}
\frac{2ab}{\pi} \nonumber \\
&& \hspace{-0.7cm} \times \int_0^{\infty} \dd u \, u^2 \, \hW_3^{(2\ell_i)}(u)
\, j_{2\ell}(a u) j_{2\ell'}(b u) .
\label{W3-ab-def}
\eeqa
Then, substituting the expansion (\ref{cIs-n}) into Eq.(\ref{Kij-3}) we obtain
\beqa
\overline{\xis_{i;j'} \xis_{i';j}} & = & 2\theta_s \int_0^{\infty}
\frac{\dd r \; r}{(\cD\theta_s)^2} \sum_{\ell,\ell'=0}^4
\cI_{i,2\ell_i}^{(s;2\ell)}(r) \, \cI_{j,2\ell_j}^{(s;2\ell')}(r) \nonumber \\
&& \times \; A^{(\ell,\ell')}\left(\!\frac{r}{\cD\theta_s}\!\right) , 
\label{Kij-3-l-lp}
\eeqa
with
\beq
A^{(\ell,\ell')}(y) = \frac{1}{2} \int_{-1}^1 \dd\mu \; P_{2\ell}(\mu) P_{2\ell'}(\mu) 
\, A^{(s)}(y,\mu^2) ,
\label{A-ell-ell'-def}
\eeq
where we used that $A^{(s)}$ defined in Eq.(\ref{A3s_r-def}) also writes as
\beqa
A^{(s)}(\vy) & \! \equiv \! & A^{(s)}(y,\mu^2) = y \int_0^{\infty} \!\! \dd u \, u \,
\tW_2(u)^2 \, J_0(y u \sqrt{1 \!-\! \mu^2}) . \nonumber \\
&&
\label{A3_y-def}
\eeqa
These expressions closely follow those obtained in \citet{Valageas2011}.
To recover their real-space results, it is sufficient to set $\beta=0$ in 
Eq.(\ref{Ps}), which makes the power spectrum isotropic so that only
monopole contributions (i.e., $\ell=0$) remain. 
Thus, in Eq.(\ref{Kij-3-l-lp}) there only remains the term $\ell=\ell'=0$,
and the function $A^{(0,0)}(y)$ can actually be computed explicitly
(see their Eqs.(F10) and (F11)).

As in \citet{Valageas2011}, we preferred to use the configuration-space
expression (\ref{Kij-3-l-lp}) rather than the Fourier-space expression
(\ref{xi-ij'-i'j-1}) for numerical computations. Indeed, the integrals
have been partially factored out in Eq.(\ref{Kij-3-l-lp}) (so that it has the same
level of complexity as a two-dimensional integral) and the oscillatory kernels
$\tW_i^{(2\ell_i)}$ of Eq.(\ref{xi-ij'-i'j-1}) have been replaced by the
slowly-varying kernels $\cW_{\ell_i}^{(\ell,\ell')}$ of Eq.(\ref{cI-3-W3}), see
Eqs.(\ref{W3-s-ell-1})-(\ref{W3-s-ell-5}).

We now turn to the contribution (\ref{Cij-zeta}) that arises from the three-point
correlation function.
The first term within the brackets in Eq.(\ref{Cij-zeta}) involves the product of
two averages $\overline{\xis_{i'}}$ and $\overline{\xis_{j'}}$, which
are given in Eq.(\ref{xi-i-i'-1}). Following the analysis of \citet{Valageas2011},
the second term can be written in configuration space as
\beq
\overline{\xis_{i',i} \xis_{i',j'}} = \int_{\cV_i} \frac{\dd\vr}{\cV_i} \, 
(4\ell_i \!+\! 1) P_{2\ell_i}(\mu_{\vr}) \xis(\vr) \, \cI_{j,2\ell_j}^{(s)}(\vr) . 
\label{J3ij-1}
\eeq
Using the expansions (\ref{xis}) and (\ref{cIs-n}), this yields
\beqa
\overline{\xis_{i',i} \xis_{i',j'}} & = & (4\ell_i \!+\! 1) \int_{\Rim}^{\Rip} 
\frac{\dd r \; 3 r^2}{\Rip^3-\Rim^3} \sum_{\ell=0}^2 \sum_{\ell'=0}^4
\xi^{(s;2\ell)}(r) \nonumber \\
&& \times \; \cI_{j,2\ell_j}^{(s;2\ell')}(r) \;
\left( \bea{ccc} 2\ell & 2\ell' & 2\ell_i \\ 0 & 0 & 0 \ea \right)^{\!\!2} .
\label{J3ij-2}
\eeqa
The third term within the brackets in Eq.(\ref{Cij-zeta}) is obtained from
Eq.(\ref{J3ij-2}) by exchanging the labels ``$i$'' and ``$j$''.

We now turn to the contribution (\ref{Cij-eta}) that arises from the four-point
correlation function, proceeding in a similar fashion.
The new geometrical average involved by the first two terms within the  
brackets in Eq.(\ref{Cij-eta}) reads as
\beqa
\overline{\xi_{i;j} \xis_{i;j'}} & = & 2\theta_s \int_0^{\infty} 
\frac{\dd r \; r}{(\cD\theta_s)^2} \; \sum_{\ell=0}^2 \sum_{\ell'=0}^4 
\xi^{(s;2\ell)}(r) \, \cI_{j,2\ell_j}^{(s;2\ell')}(r) \nonumber \\
&& \times \; A^{(\ell,\ell')} \!\left(\frac{r}{\cD\theta_s}\right) ,
\label{Kj-def}
\eeqa
which does not depend on the bin $i$.
The third term involves the cylindrical average
\beq
\xicyl = \pi  \int_0^{\infty} \frac{\dd k}{k} \frac{\Delta^2(k)}{\cD k}
\tW_2(k\cD\theta_s)^2 .
\label{I-thetas-def}
\eeq
The fourth term, which does not factor, gives rise to
\beqa
\overline{\xis_{j';i}\xi_{i;j}\xis_{j;i'}} & = & 2\theta_s \int_0^{\infty} 
\frac{\dd r \; r}{(\cD\theta_s)^2} \sum_{\ell=0}^2 \; \sum_{\ell'\!,\ell''=0}^4 
\xi^{(s;2\ell)}(r) \nonumber \\
&& \hspace{-0.7cm} \times \; \cI_{i,2\ell_i}^{(s;2\ell')}(r) \,
\cI_{j,2\ell_j}^{(s;2\ell'')}(r) \, 
A^{(\ell,\ell'\!,\ell'')} \!\left(\frac{r}{\cD\theta_s}\right) ,
\label{Lij-def}
\eeqa
where we introduced, as in Eq.(\ref{A-ell-ell'-def}),
\beqa
A^{(\ell,\ell'\!,\ell'')}(y) \! & = & \frac{1}{2} \int_{-1}^1 \!\! \dd\mu \,
P_{2\ell}(\mu) P_{2\ell'}(\mu) P_{2\ell''}(\mu) A^{(s)}(y,\mu^2) . \nonumber\\
&&
\label{A-ell-ell'-ell''-def}
\eeqa
The last two terms within the brackets in Eq.(\ref{Cij-eta}) involve the
geometrical average
\beqa
\overline{\xis_{j';i}\xis_{i,i'}\xis_{i';j}} & = & \frac{\theta_s}{2\pi}
\int\frac{\dd\vr}{r(\cD\theta_s)^2} \int_{\cV_i} \frac{\dd\vr'}{\cV_i}
\, (4\ell_i \!+\! 1) P_{2\ell_i}(\mu_{\vr'}) \nonumber \\
&& \hspace{-1cm} \times \; \cI_{j,2\ell_j}^{(s)}(\vr) \, \xis(\vr') \, 
\xis(\vr+\vr') \, A^{(s)}\!\left(\!\frac{\vr}{\cD\theta_s}\!\right) .
\eeqa
To integrate over angles, we again introduce a Dirac factor
$\int\dd\vr''\delta_D(\vr''-\vr-\vr')$ that we write under its exponential 
representation, which we expand over spherical harmonics.
Using the expansions (\ref{xis}) and (\ref{cIs-n}), we obtain
\beqa
\overline{\xis_{j';i}\xis_{i,i'}\xis_{i';j}} & \! = \! & \theta_s \int_0^{\infty}
\!\!\! \frac{\dd r \; r}{(\cD\theta_s)^2} \int_{\Rim}^{\Rip} \!\!
\frac{\dd r' \; 3 r'^2}{\Rip^3 \!-\! \Rim^3} \int_{|r-r'|}^{r+r'} \! 
\frac{\dd r'' r''}{r \, r'}
\nonumber \\ 
&& \hspace{-1.5cm} \times \sum_{\ell=0}^4 \; \sum_{\ell'\!,\ell''=0}^2 
\cI_{j,2\ell_j}^{(s;2\ell)}(r) \, \xi^{(s;2\ell')}(r') \, \xi^{(s;2\ell'')}(r'') 
\sum_{n'=|\ell_i-\ell'|}^{\ell_i+\ell'} \nonumber \\
&& \hspace{-1.5cm} \times \! \sum_{n=|n'-\ell''|}^{n'+\ell''}  \!
(-1)^{n+n'+\ell''} (4\ell_i \!+\! 1) (4n \!+\! 1) (4n' \!+\! 1) \nonumber \\
&& \hspace{-1.5cm} \times 
\left( \bea{ccc} 2n' & 2\ell_i & 2\ell' \\ 0 & 0 & 0 \ea \right)^{\!\!2} 
\left( \bea{ccc} 2n & 2n' & 2\ell'' \\ 0 & 0 & 0 \ea \right)^{\!\!2} 
A^{(\ell,n)}\left(\frac{r}{\cD\theta_s}\right) \nonumber \\
&& \hspace{-1.5cm} \times \; C^{n,n',\ell''}(r,r',r'') ,
\label{Tij-2}
\eeqa
where we introduced the symmetric functions
\beqa
C^{\ell_1,\ell_2,\ell_3}(r_1,r_2,r_3) & = & \frac{4}{\pi} \, r_1 r_2 r_3
\int_0^{\infty} \!\! \dd k \; k^2 \, j_{2\ell_1}(k r_1)  \nonumber \\
&& \times \;  j_{2\ell_2}(k r_2)  \, j_{2\ell_3}(k r_3) .
\label{C3-l-r-def}
\eeqa
For numerical computations it is more efficient to write Eq.(\ref{Tij-2})
in a partially factorized form by exchanging the order of the integrations
and moving the integration over $k$ included in the function
$C^{\ell_1,\ell_2,\ell_3}$ to the left. Using the decomposition (\ref{xis})-(\ref{xis-4})
and integrating over $r''$, this gives
\beqa
\overline{\xis_{j';i}\xis_{i,i'}\xis_{i';j}} & \! = \! & 2\theta_s \! \int_0^{\infty}
\frac{\dd k}{k} \Delta^2(k) \sum_{\ell=0}^4 \; 
\sum_{\ell'\!,\ell''=0}^2 \sum_{n'=|\ell_i-\ell'|}^{\ell_i+\ell'}
\sum_{n=|n'-\ell''|}^{n'+\ell''}  \nonumber \\
&& \hspace{-1.5cm} \times (-1)^{n+n'+\ell''} (4\ell_i \!+\! 1) (4n \!+\! 1) 
(4n' \!+\! 1) 
\left( \bea{ccc} 2n' & 2\ell_i & 2\ell' \\ 0 & 0 & 0 \ea \right)^{\!\!2} 
\nonumber \\
&& \hspace{-1.5cm} \times 
\left( \bea{ccc} 2n & 2n' & 2\ell'' \\ 0 & 0 & 0 \ea \right)^{\!\!2} 
F^{(2\ell'')} \int_0^{\infty} \!\! \frac{\dd r \; r}{(\cD\theta_s)^2}
j_{2n}(k r) \cI_{j,2\ell_j}^{(s;2\ell)}(r) \nonumber \\
&& \hspace{-1.5cm} \times A^{(\ell,n)}\left(\frac{r}{\cD\theta_s}\right)
\int_{\Rim}^{\Rip} \!\! \frac{\dd r' \; 3 r'^2}{\Rip^3 \!-\! \Rim^3}
j_{2n'}(k r') \xi^{(s;2\ell')}(r') , \nonumber \\
&& \hspace{-1.5cm} 
\label{Tij-3}
\eeqa
where $F^{(2\ell'')}$ is the prefactor in Eqs.(\ref{xis-0})-(\ref{xis-4}),
\beq
F^{(0)}= 1+\frac{2\beta}{3}+\frac{\beta^2}{5} , \;
F^{(2)}= - \frac{4\beta}{3}-\frac{4\beta^2}{7} , \;
F^{(4)}= \frac{8\beta^2}{35} .
\eeq

Again, all these geometrical averages of redshift-space correlations
reduce to the real-space results of \citet{Valageas2011}, if we set
$\beta=0$ and only keep the monopole terms.

\section{Some useful integrals}
\label{integrals}

We describe here the computation of some of the functions introduced
in Appendix~\ref{geometrical-averages}.

\subsection{Multipoles of $A^{(s)}(\vy)$}
\label{cal-As}

Using the expression (\ref{A3_y-def}) and the expansion
\citep{Gradshteyn1994}
\beqa
J_0(z\sqrt{1-\mu^2}) & = & \sqrt{\frac{2\pi}{z}} \sum_{n=0}^{\infty} 
(2n+1/2)  \frac{(2n-1)!!}{2^n n!} \nonumber \\
&& \times \; J_{2n+1/2}(z) P_{2n}(\mu) ,
\eeqa
we can write
\beq
A^{(s)}(y,\mu^2) = \sum_{n=0}^{\infty} (4n \!+\! 1) \frac{(2n \!-\! 1)!!}{2^n n!}
B_n(y) P_{2n}(\mu) ,
\eeq
with
\beq
B_n(y) = y \int_0^{\infty} \dd u \, u \, \tW_2(u)^2 \, j_{2n}(y u) .
\label{B-def}
\eeq
This yields the multipoles
\beqa
A^{(\ell_1,\ell_2)}(y) & = & \sum_{n=|\ell_1-\ell_2|}^{\ell_1+\ell_2} 
(4n \!+\! 1) \frac{(2n \!-\! 1)!!}{2^n n!}
\left( \bea{ccc} 2\ell_1 & 2\ell_2 & 2n \\ 0 & 0 & 0 \ea \right)^{\!\!2}  
\nonumber \\
&& \times \; B_n(y) ,
\label{A2-B}
\eeqa
and
\beq
A^{(\ell_1,\ell_2,\ell_3)}(y) \!\! = \!\! \sum_{n=0}^{\ell_1+\ell_2+\ell_3} 
(4n \!+\! 1) \frac{(2n \!-\! 1)!!}{2^n n!} Y^{\ell_1,\ell_2,\ell_3,n} B_n(y) ,
\label{A-Y-B}
\eeq
with
\beq
Y^{\ell_1,\ell_2,\ell_3,\ell_4} =  \frac{1}{2} \int_{-1}^1 \dd\mu \, 
P_{2\ell_1}(\mu) P_{2\ell_2}(\mu) P_{2\ell_3}(\mu) P_{2\ell_4}(\mu) .
\label{Y-4-def}
\eeq
Because we only considered the low-order multipoles $0\leq\ell_i\leq 2$
of the redshift-space correlation function, we only need the first few
multipoles $A^{(\ell_1,\ell_2)}$ and $A^{(\ell_1,\ell_2,\ell_3)}$.
The associated numbers $Y^{\ell_1,\ell_2,\ell_3,\ell_4}$ and functions
$B_n(y)$ can be computed in advance.

\subsection{Symmetric functions $C^{\ell_1,\ell_2,\ell_3}$}
\label{Cal-C}

The functions $C^{\ell_1,\ell_2,\ell_3}$ defined in Eq.(\ref{C3-l-r-def})
vanish if the three lengths $\{r_1,r_2,r_3\}$ do not obey triangular inequalities.
It is also useful to note that these functions obey the scale invariance
\beq
\lambda > 0 : \;\; 
C^{\ell_1,\ell_2,\ell_3}(\lambda r_1,\lambda r_2,\lambda r_3) = 
C^{\ell_1,\ell_2,\ell_3}(r_1,r_2,r_3) .
\label{C-scale}
\eeq
For $|r_1-r_2|<r_3<r_1+r_2$, as in Eq.(\ref{Tij-2}), we obtain explicit
expressions using the recursion relations of Bessel functions and the property \citep{Gradshteyn1994}
\beqa
\int_0^{\infty} \dd k \, k^{1-\alpha} J_{\alpha}(a k) J_{\beta}(b k) 
J_{\beta}(c k) & = &\frac{(bc)^{\alpha-1}}{\sqrt{2\pi}a^{\alpha}} \,
(\sin v)^{\alpha-1/2} \nonumber \\
&& \hspace{-1cm} \times \; P_{\beta-1/2}^{1/2-\alpha}(\cos v) ,
\label{C-P}
\eeqa
for
\beq
|a-b| < c < a+b , \;\;\; 2bc \, \cos(v) = b^2+c^2-a^2 .
\eeq
Indeed, Eq.(\ref{C3-l-r-def}) also reads as
\beqa
C^{\ell_1,\ell_2,\ell_3}(r_1,r_2,r_3) & = & \sqrt{2\pi r_1 r_2 r_3}
\int_0^{\infty} \!\! \dd k \; \sqrt{k} \, J_{2\ell_1+1/2}(k r_1)  \nonumber \\
&& \times \;  J_{2\ell_2+1/2}(k r_2)  \, J_{2\ell_3+1/2}(k r_3) .
\label{C3-l-r-J}
\eeqa
Then, using the recursion
\beq
J_{2\ell+1/2}(z)= \frac{4\ell \!-\! 1}{z}  J_{2\ell-1/2}(z) - 
J_{2\ell-3/2}(z) ,
\eeq
we can lower the indices $\{\ell_1,\ell_2,\ell_3\}$ and can express
all required quantities in terms of integrals of the form (\ref{C-P}) with
two identical indices and the right power-law prefactor.
For instance, the first few functions read as
\beq
C^{0,0,0} = 1 ,
\label{C-000}
\eeq
\beq
C^{0,1,1} = \frac{3\mu_1^2-1}{2} ,
\eeq
\beq
C^{1,1,1} = \frac{3}{2} \frac{r_2 r_3}{r_1^2} \mu_1 (1-\mu_1^2) 
- \frac{3\mu_1^2-1}{2} ,
\eeq
were we defined
\beq
\mu_1 = \frac{r_2^2+r_3^2-r_1^2}{2r_2r_3} .
\label{mu1-def}
\eeq

\subsection{Functions $\cW_{\ell_i}^{\ell,\ell'}(a,b)$}
\label{Cal-W}

To compute the geometrical means $\cI_{i,2\ell_i}^{(s)}(\vr)$ of the
redshift-space correlation with respect to bin $i$, defined in Eq.(\ref{cIs-def}),
we need the functions $\cW_{\ell_i}^{\ell,\ell'}$ defined in
Eq.(\ref{W3-ab-def}).
Using Eq.(\ref{Whi-def}), they also write as
\beqa
\cW_{\ell_i}^{(\ell,\ell')}(a,b) & = & (-1)^{\ell+\ell'+\ell_i} (4\ell_i+1) 
\left( \bea{ccc} 2\ell & 2\ell' & 2\ell_i  \\ 0 & 0 & 0 \ea \right)^{\!\!2}
\nonumber \\
&& \times \int_0^{1} \dd r \; \frac{3r}{2} \; C^{\ell,\ell',\ell_i}(a,b,r) ,
\label{W3-C3}
\eeqa
where the functions $C^{\ell_1,\ell_2,\ell_3}$ were defined
in Eq.(\ref{C3-l-r-def}). 
From the triangular constraint on these functions $C^{\ell_1,\ell_2,\ell_3}$ 
we obtain
\beq
|a-b|>1 : \;\; \cW_{\ell_i}^{(\ell,\ell')}= 0 ,
\label{W3-s-ell-1}
\eeq
while the scale invariance (\ref{C-scale}) leads to
\beqa
a+b \leq 1 & : & \cW_{\ell_i}^{(\ell,\ell')}(a,b) = (a+b)^2 \;
\cW_{\ell_i}^{(\ell,\ell')}\left(\frac{a}{a+b},\frac{b}{a+b}\right) .
\nonumber \\
&&
\eeqa
Then, from the explicit expressions obtained
in Appendix~\ref{Cal-C}, which are rational functions of $\{r_1,r_2,r_3\}$, 
we can perform the integral (\ref{W3-C3}) and derive explicit expressions.
In particular, the first function reads as
\beq
|a\!-\!b|<1, \; a\!+\!b<1 : \;\;\; \cW_0^{(0,0)}= 3 a b ,
\eeq
\beq
|a-b|<1, \; a+b>1 : \;\;\; \cW_0^{(0,0)}= \frac{3}{4} [1-(a-b)^2] .
\label{W3-s-ell-5}
\eeq

\section{Correlation matrices of the ``$2\ell=2$'' redshift-space correlation}
\label{Correlation-matrix-l=2}

\subsection{Auto-correlation of the ``$2\ell=2$'' multipole}
\label{auto_l=2}

\begin{figure}
\begin{center}
\epsfxsize=5.9 cm \epsfysize=5.9 cm {\epsfbox{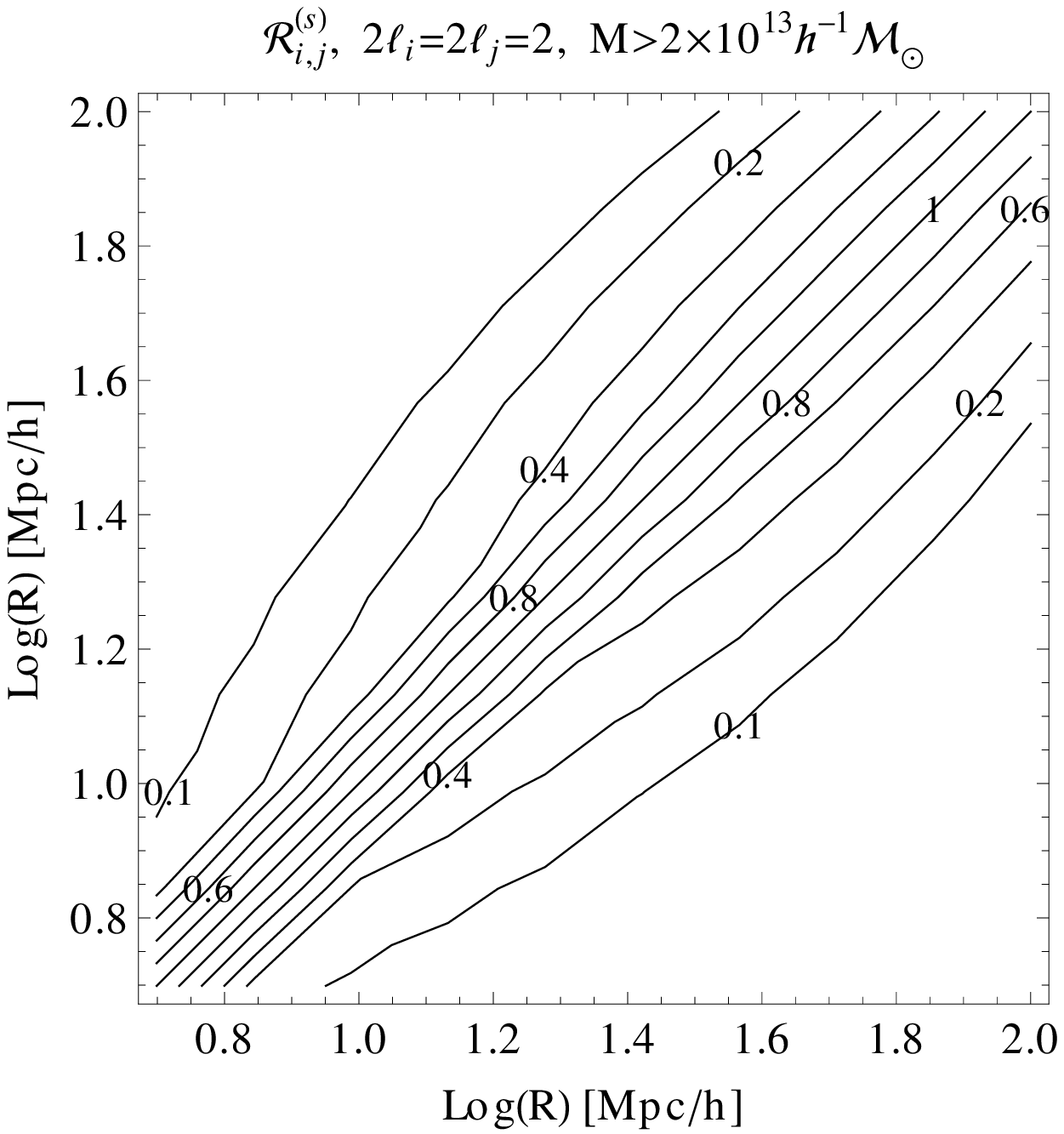}}
\epsfxsize=5.9 cm \epsfysize=5.9 cm {\epsfbox{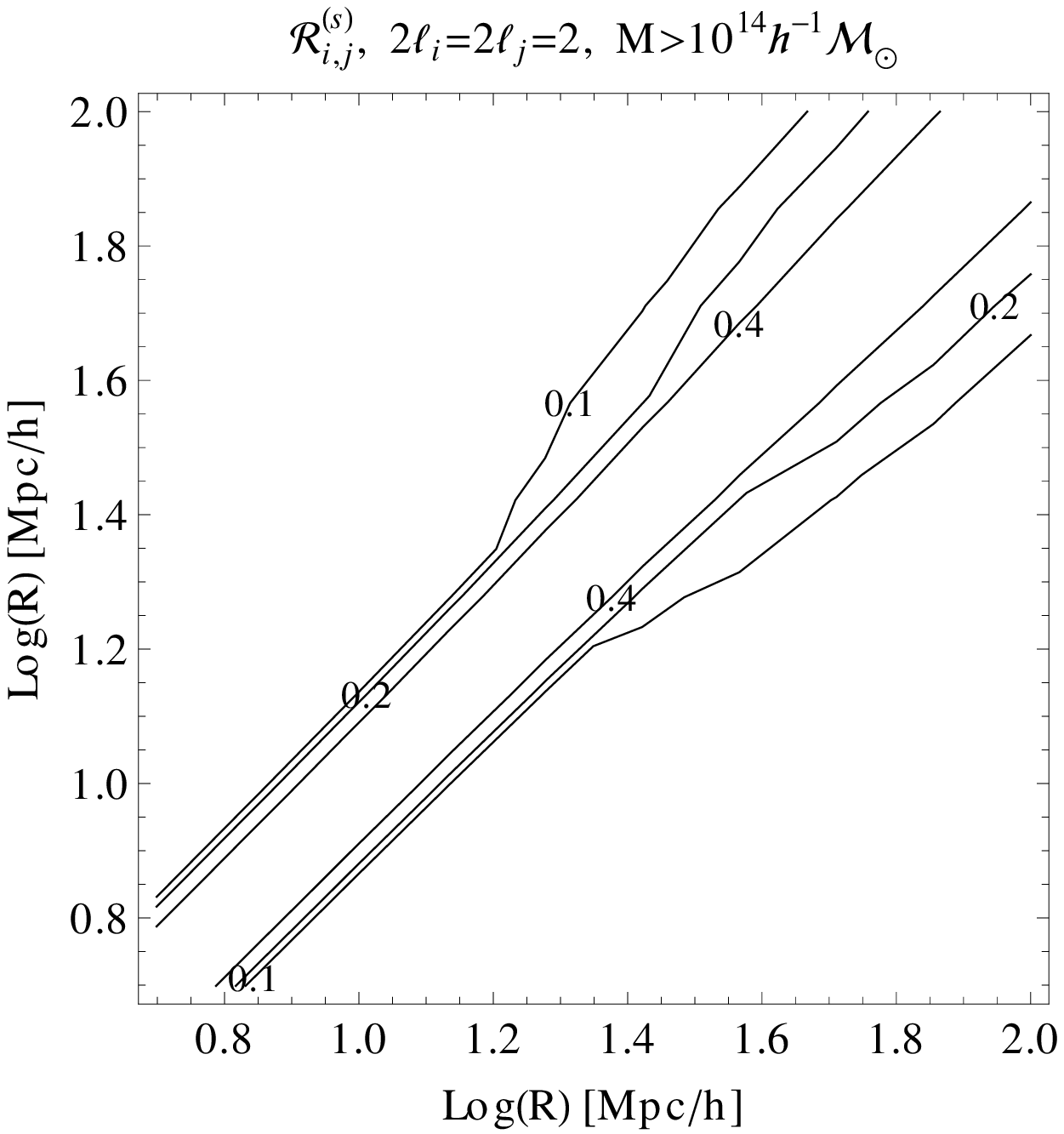}}
\end{center}
\caption{Contour plots for the redshift-space auto-correlation matrices
$\cR^{(s)}_{i,j}$, defined as in Eq.(\ref{R-ij-def}), for the 
``$2\ell=2$'' redshift-space correlation studied in Sect.~\ref{Covariance-2l=2}.
We considered the same cases (distance bins, redshift interval, and mass
thresholds) as in Fig.~\ref{fig_Rij_CXis}, where we plotted the auto-correlation
matrices of the monopole term of the redshift-space correlation.}
\label{fig_Rij_CXis_l1l1}
\end{figure}

We show in Fig.~\ref{fig_Rij_CXis_l1l1} the auto-correlation matrices of the
``$2\ell=2$'' redshift-space correlation. For simplicity we only plot our results
for the full matrices that include the non-Gaussian contributions associated with
the 3-pt and 4-pt correlation functions.
We can see that they are significantly more diagonal than the auto-correlation 
matrices of the monopole term that were displayed in the right panels of
Fig.~\ref{fig_Rij_CXis}.

\subsection{Cross-correlation of the ``$2\ell=0$'' and ``$2\ell=2$'' multipoles}
\label{cross_l=0_l=2}

\begin{figure}
\begin{center}
\epsfxsize=5.9 cm \epsfysize=5.9 cm {\epsfbox{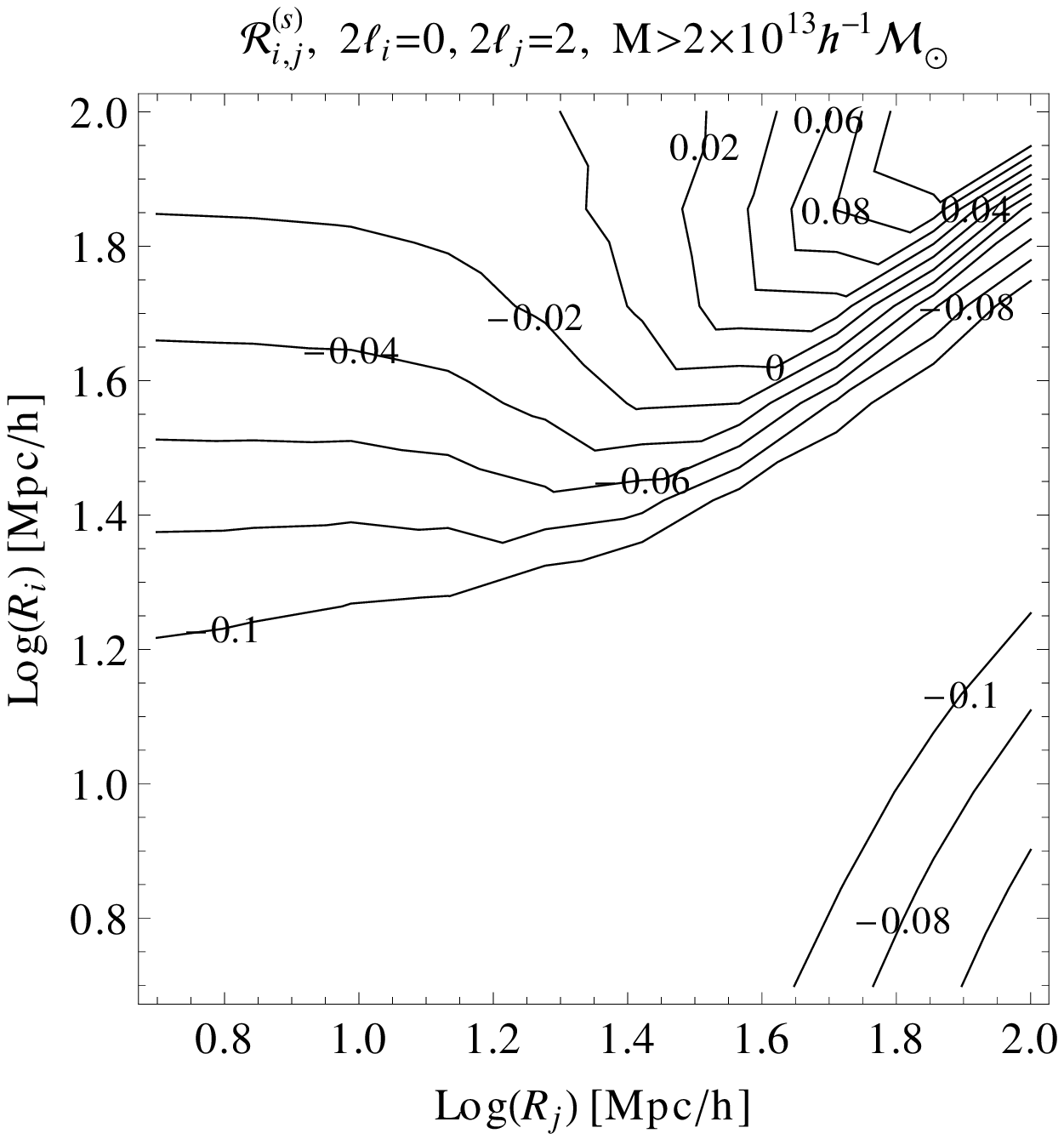}}
\epsfxsize=5.9 cm \epsfysize=5.9 cm {\epsfbox{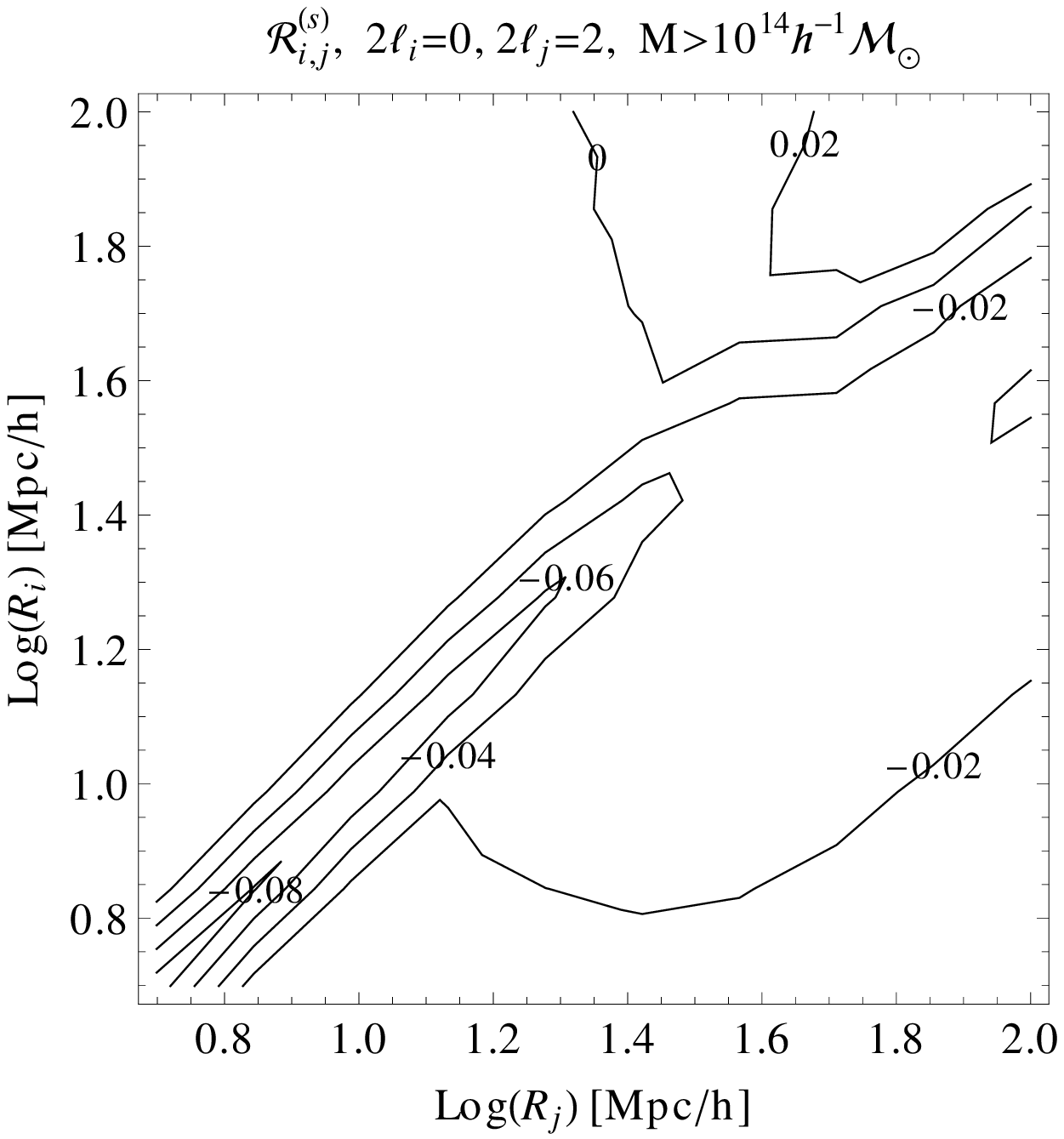}}
\end{center}
\caption{Contour plots for the redshift-space cross-correlation matrices
$\cR^{(s)}_{i,j}$, defined as in Eq.(\ref{Rij-0-2}), for the 
``$2\ell_i=0$'' and ``$2\ell_j=2$'' redshift-space correlations.
We considered the same cases (distance bins, redshift interval, and mass
thresholds) as in Figs.~\ref{fig_Rij_CXis} and \ref{fig_Rij_CXis_l1l1}.}
\label{fig_Rij_CXis_l0l1}
\end{figure}

We show in Fig.~\ref{fig_Rij_CXis_l0l1} the cross-correlation matrices of the
``$2\ell_i=0$'' and ``$2\ell_j=2$'' redshift-space correlations.
Thus, in the plane $\{R_i,R_j\}$ we plot the contour plots of the matrix
\beq
\cR^{(s)}_{R_i,R_j} = \frac{C^{(s)}_{R_i,2\ell_i=0;R_j,2\ell_j=2}}
{\sqrt{C^{(s)}_{R_i,0;R_i,0} C^{(s)}_{R_j,2;R_j,2}}} .
\label{Rij-0-2}
\eeq
Because $\ell_i \neq \ell_j$ this matrix is no longer symmetric in the
plane $\{R_i,R_j\}$ and it can be negative for $R_i=R_j$.
We can see that its amplitude is rather small, typically on the order of $0.05$,
even along the diagonal.
Indeed, because $\ell_i \neq \ell_j$ the pure shot-noise contribution to
Eq.(\ref{Cij-G}) vanishes (the factor $\delta_{\ell_i,\ell_j} (4\ell_i+1)$ within
the brackets).
This is due to the orthogonality of the Legendre polynomials, which are the
basis of the multipole decomposition of the redshift-space correlation.
The small amplitude of $\{R_i,R_j\}$ found in Fig.~\ref{fig_Rij_CXis_l0l1}
means that the full covariance matrix $C^{(s)}_{i,j}$ is approximately
block-diagonal in the space $\{R_i,2\ell_i\}$. Therefore, we may neglect
$C^{(s)}_{i,j}$ for $\ell_i\neq\ell_j$ and decouple the analysis of
$\xi^{(s)}_{R_i,0}$ and $\xi^{(s)}_{R_i,2}$.

\bibliographystyle{aa} 
\bibliography{ref1}

\end{document}